\documentclass[final,5p,times,twocolumn]{elsarticle}
\usepackage{amssymb}
\usepackage{amsmath}
\usepackage[en-DE]{datetime2} % Formats dates/times for Germany / Central Europe

\usepackage{lineno}
\usepackage[justification=Centering]{caption}
\usepackage[justification=Centering]{subcaption}
\usepackage{rotating}
\usepackage{graphicx}% Include figure files
\usepackage{color}
\usepackage[utf8]{inputenc}
\usepackage[T1]{fontenc}
\usepackage[colorlinks=true]{hyperref}

\begin{document}

\begin{frontmatter}

%% Title, authors and addresses

%% use the tnoteref command within \title for footnotes;
%% use the tnotetext command for theassociated footnote;
%% use the fnref command within \author or \affiliation for footnotes;
%% use the fntext command for theassociated footnote;
%% use the corref command within \author for corresponding author footnotes;
%% use the cortext command for theassociated footnote;
%% use the ead command for the email address,
%% and the form \ead[url] for the home page:
%% \title{Title\tnoteref{label1}}
%% \tnotetext[label1]{}
%% \author{Name\corref{cor1}\fnref{label2}}
%% \ead{email address}
%% \ead[url]{home page}
%% \fntext[label2]{}
%% \cortext[cor1]{}
%% \affiliation{organization={},
%%             addressline={},
%%             city={},
%%             postcode={},
%%             state={},
%%             country={}}
%% \fntext[label3]{}

\title{CBM T$_0$ Detector Performance Study with Light Ions ($^4$He, $^{12}$C) and the DOGMA Readout System}

%% use optional labels to link authors explicitly to addresses:
%% \author[label1,label2]{}
%% \affiliation[label1]{organization={},
%%             addressline={},
%%             city={},
%%             postcode={},
%%             state={},
%%             country={}}
%%
%% \affiliation[label2]{organization={},
%%             addressline={},
%%             city={},
%%             postcode={},
%%             state={},
%%             country={}}

\author[d]{Yevhen Kozymka \corref{cor1}} 
\author[c]{Thomas Bergauer}
\author[a]{Maximilian von Buelow}
\author[d]{Henrik Flörsheimer} 
\author[a]{Jochen Frühauf} 
\author[a,d,e]{Tetyana Galatyuk} 
\author[c,f]{Harald Handerkas} 
\author[a]{Henning Heggen} 
\author[a]{Mladen Kis} 
\author[a]{Sergey Linev}  
\author[a]{Jerzy Pietraszko} 
\author[a]{Christian Joachim Schmidt}
\author[a]{Luca Schramm} 
\author[a]{Michael Traxler}
\author[a]{Michael Träger} 
\author[c,f]{Felix Ulrich-Pur} 
\author[a]{and Robert Visinka}

\affiliation[a]{organization={GSI Helmholtzzentrum für Schwerionenforschung GmbH}, city={Darmstadt}, country={Germany}}
\affiliation[c]{organization={Austrian Academy of Sciences, Marietta-Blau-Institut for Particle Physics}, city={Vienna}, country={Austria}}
\affiliation[d]{organization={Technische Universität Darmstadt}, city={Darmstadt}, country={Germany}}
\affiliation[e]{organization={Helmholtz Forschungsakademie Hessen für FAIR}}
\affiliation[f]{organization={TU Wien, Atominstitut}, city={Vienna}, country={Austria}}
% \affiliation[g]{organization={EBG MedAustron}, city={Wiener Neustadt}, country={Austria}}

\cortext[cor1]{Corresponding author}

%% Abstract
\begin{abstract}
%% Text of abstract
%%\textbf{Version: Date: \today, UTC Time (add 2h to get DE Time): \DTMcurrenttime} \\ \\
This paper presents the realization and performance of the CBM T$_0$ detector system. Built upon an approximately 70~\textmu m thin polycrystalline CVD (pcCVD) diamond sensor with pad metallization, the detector is mounted on a flexible vacuum manipulator and operates entirely under passive cooling. Interfaced with the DOGMA readout architecture, the system successfully meets all operational requirements for CBM. It demonstrates a high detection efficiency (near 100\%) and a timing precision below $50\,\text{ps}$ RMS across a broad range of beam species, from light ions (such as Carbon) to heavy ions (such as Gold). The DOGMA readout features two distinct data paths: a high-precision TDC path routing data directly to the global CBM DAQ, and a secondary path delivering real-time rate measurements crucial for online beam quality monitoring and fast beam abort logic. Finally, detailed performance studies conducted with light-ion beams at EBG MedAustron are discussed.
\end{abstract}

%%Graphical abstract
%%\begin{graphicalabstract}
%\includegraphics{grabs}
%%\end{graphicalabstract}

%%Research highlights
%%\begin{highlights}
%%\item Research highlight 1
%%\item Research highlight 2
%%\end{highlights}

%% Keywords
\begin{keyword}
%% keywords here, in the form: keyword \sep keyword
    Timing Detectors \sep Diamond Detectors \sep Particle Tracking Detectors (Solid-State Detectors) \sep Radiation Damage to Detector Materials (Solid-State)
%% PACS codes here, in the form: \PACS code \sep code

%% MSC codes here, in the form: \MSC code \sep code
%% or \MSC[2008] code \sep code (2000 is the default)

\end{keyword}

\end{frontmatter}

\twocolumn

\section{Introduction}
    \label{intro}
    The Compressed Baryonic Matter (CBM) experiment at the Facility for Antiproton and Ion Research (FAIR, Darmstadt, Germany) is currently under construction, with a focus on high-precision investigations of the Quantum Chromodynamics (QCD) phase diagram—particularly in the region of high net-baryon density~\cite{Agarwal2023}. A variety of observables, including collective particle flow, event-by-event fluctuations, dilepton pair production and strangeness production, serve as sensitive probes to explore the fundamental properties of dense nuclear matter in heavy-ion collisions~\cite{CBMbook2011}. Furthermore, collision studies of lighter ion systems provide crucial baseline measurements and system-size dependence trends necessary for a proper interpretation of heavy-ion collision dynamics.
    
    This challenging experimental program requires high-rate and high-precision measurements, demanding high-purity Particle Identification (PID) capabilities across the detector setup. A crucial component of this system is the T$_0$ reaction detector, positioned directly in the primary beam line upstream of the target. To meet the experiment's requirements, the T$_0$ detector must satisfy several design criteria: (1) Precise T$_0$ determination with a time precision below 50~ps RMS at particle rates up to 10$^7$ ions/s, for a wide range of ion species ranging from light ions ($^{12}$C) to heavy ions ($^{197}$Au); (2) a minimal material budget to reduce unwanted beam interactions upstream of the target; (3) sufficient granularity for precise beam position monitoring and detection of beam intensity fluctuations; (4) radiation-hard sensor technology; (5) in-vacuum operation with passive cooling to minimize additional material in front of the target; and (6) an active area of approximately 1 cm $\times$ 1 cm to cover the full beam profile delivered to the CBM target station.
    
    Several candidate sensor technologies have been extensively evaluated for this purpose, including Single-Crystal (sc) and Poly-Crystalline (pc) Chemical Vapour Deposition (CVD) diamond, Low Gain Avalanche Diode (LGAD) detectors and conventional silicon sensors. Benefiting from extensive operational experience gained in the HADES experiment with these detector materials\cite{hades2009, pieMips2010, pieAu2014} and considering the stringent requirements of the CBM T$_0$ detector, pcCVD diamond was selected as the sensor material for this application. This decision was primarily driven by: (1) stable operation at room temperature using passive cooling; (2) exceptionally fast signal rise times; (3) superior radiation hardness, specifically the negligible increase in dark current under high particle fluence compared to silicon-based technologies; and (4) commercial availability, as single pcCVD sensors with active areas of 1 cm $\times$ 1 cm can be reliably produced at a commercial scale.
    
    Radiation damage is a major concern for any detector operated in ion beams and diamond sensors suffer performance degradation after irradiation~\cite{pieAu2014,pie2025}. Diamond detectors exhibit a gradual loss of Charge Collection Efficiency (CCE) with increasing particle fluence, mainly due to radiation-induced defects acting as charge-trapping centers in the crystal lattice. The CCE can be expressed in terms of the Charge Collection Distance (CCD) and sensor thickness d as $\text{CCE} = \text{CCD} / \text{d}$. Extensive radiation-damage studies of single-crystalline (scCVD) and polycrystalline (pcCVD) diamond, including irradiation with proton beams, have shown that the degradation of the charge collection properties can be described by a simple damage model. Moreover, the radiation-induced damage in both material types can be characterized using the same damage constant~\cite{baeni2019}. In contrast to scCVD diamond, pcCVD material contains intrinsic grains and grain boundaries that introduce additional charge-trapping sites already before irradiation, resulting in a lower initial CCD and consequently a lower CCE. 
    For the non-irradiated pcCVD material investigated in Ref.~\cite{baeni2019}, the measured CCD corresponds to a CCE of approximately $44-47$~\%.
    
    In the highly demanding CBM T$_0$ application, where high-intensity beams must be measured and signal deterioration due to radiation damage is unavoidable, a dedicated amplification system has been proposed to compensate for radiation-induced sensor degradation~\cite{pie2025}. As shown, this approach can significantly extend the operational lifetime of the sensor. The radiation damage compensation amplification system is based on two stages of analog amplification and incorporates an AC-coupling circuit at the amplifier input, which blocks DC components and low-frequency distortions originating from the sensor before they reach the Front-End Electronics (FEE).
    
    In the following sections, the sensor geometry optimization, readout system, vacuum-compatible mechanical design, and detailed performance studies with light ions are presented.

\section{Pad vs. Strip Readout Geometry for the CBM T$_0$ Detector}
    \label{metall}

    \begin{figure}[b!]
        \centering
        \begin{subfigure}[c]{0.45\columnwidth}
        \centering
        \caption{\label{fig:strip_diamond}}
        \includegraphics[height = 4 cm]{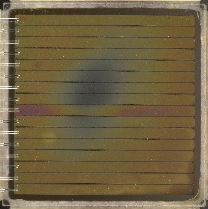}
        \end{subfigure}
        %\hfill
        \begin{subfigure}[c]{0.48\columnwidth}
        \centering
        \caption{\label{fig:pad_diamond}}
        \includegraphics[height = 4 cm]{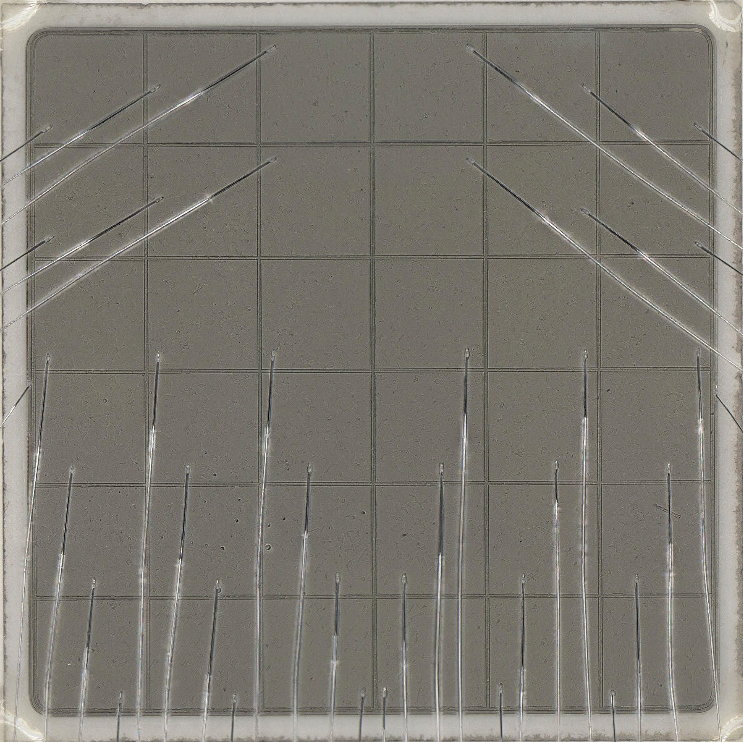}
        \end{subfigure}
        \caption{70~\textmu m thick, $1\times 1$~cm$^2$ diamond sensor with its original Cr/Au strip metallization (left) and after applying the new Al pad metallization (right). The pad numbering starts in the bottom left corner and increases in rows, so that bottom right is pad \#6, top left is pad \#31 and top right is pad \#36. Pad \#21 corresponds to the most heavily irradiated region, followed by pad \#15.}
        \label{fig:sensor_geometries}
    \end{figure}

    Two geometry options were evaluated for the sensor metallization: a strip readout and a pad readout. The strip-readout version was used in the mini-CBM (mCBM) experiment where it sustained heavy radiation damage and was subsequently tested using light ions, with the results having already been presented in~\cite{pie2025}. However, the specific radiation environment expected in the CBM experiment motivated the investigation of an alternative pad-readout geometry. In the CBM T$_0$ detector, the particle flux is highly non-uniform, resulting in strongly localized radiation damage in the central region of the sensor, while the outer regions remain significantly less affected. Consequently, the sensor response becomes position dependent, with substantially reduced signal amplitudes in the irradiated area.
    
    In a strip-readout configuration, the bonding scheme is relatively simple, allowing the use of short and uniform bond wires. However, each strip extends over regions with different radiation exposure and is connected to a single discriminator channel. As a result, one common discriminator threshold must be applied along the entire strip despite the large variation in signal amplitude. If the threshold is optimized for the non-irradiated regions, the detection efficiency in the heavily irradiated central region is significantly reduced. Conversely, lowering the threshold to maintain high efficiency in the radiation-damaged area results in excessive signal amplitudes from the non-irradiated regions, leading to increased charge sharing and unnecessarily large cluster sizes.
    
    In contrast, the pad-readout concept divides the sensor into smaller independent regions, each exposed to a more uniform radiation dose. This allows the discriminator threshold of each readout channel to be optimized according to the local sensor response, ensuring high detection efficiency while maintaining controlled cluster sizes across the detector area. Therefore, pad metallization is better suited for the non-uniform irradiation conditions expected in the CBM T$_0$ detector.
    
    To evaluate the pad-readout concept, we used the same 70-µm-thick pcCVD diamond sensor as in~\cite{pie2025}. Instead of the original 16 Chromium/Gold (Cr/Au) readout strips on each side, the sensor was re-metallized with a new pad structure. A $6\times6$ array of Aluminium (Al) readout pads was fabricated on one side of the sensor, while the opposite side was fully covered by a single Al electrode used for sensor biasing. This configuration resulted in 36 independent readout channels, each with an active pad area of approximately $1.6\times1.6$~mm$^2$. The original strip and the new pad metallization layouts, together with their corresponding bonding schemes, are shown in Fig.~\ref{fig:sensor_geometries}. The sensor re-metallization and subsequent wire-bonding processes were carried out at the GSI Detector Laboratory.
    
    The main challenge of the pad configuration is the more complex bonding scheme. Due to the increased number and different positions of the readout pads, different bond-wire lengths are required, which may introduce additional parasitic effects and potentially affect the detector performance. To evaluate the impact of the pad-bonding configuration and validate the performance of the pad-based solution, dedicated tests were carried out, as described in the following sections.

\section{DOGMA Readout System}
    \label{dogma}

    DOGMA is a modular and scalable data acquisition framework developed by the Experiment Electronics department at GSI (EE-GSI)~\cite{dogma}. It is designed to support a wide range of applications, from small detector prototypes to large-scale experimental setups. The system architecture is based, whenever possible, on commercially available components, such as Field-Programmable Gate Arrays (FPGAs), and uses standard Ethernet protocols and network infrastructure for data transmission and slow control. The communication protocol enables straightforward scaling of the system without introducing additional dead time or limiting the trigger rate, as it avoids handshake latency between individual system components~\cite{dogma}. To ensure reliable operation in radiation environments, where electronic components may be affected by Single Event Upsets (SEUs), DOGMA implements a protocol-level fast reintegration mechanism. This allows malfunctioning components to recover within approximately 100 ms after fault detection without stopping the Data Acquisition (DAQ) or affecting other components in the system. The same mechanism also enables individual components to be hot-swapped without interrupting the operation of the overall system.
    
\begin{figure}[htb]
      \centering
      \hspace*{1mm}
      \begin{minipage}[t]{0.5\textwidth}
        \centering
        %\hspace*{0mm} %% shift all to the right
        %\begin{picture}(0.55\textwidth,185)(0,0)
        \begin{picture}(\columnwidth,115)(0,0)
          \put(0,0){\includegraphics[width=\columnwidth, trim= 2cm  5cm 0 1cm, clip ]{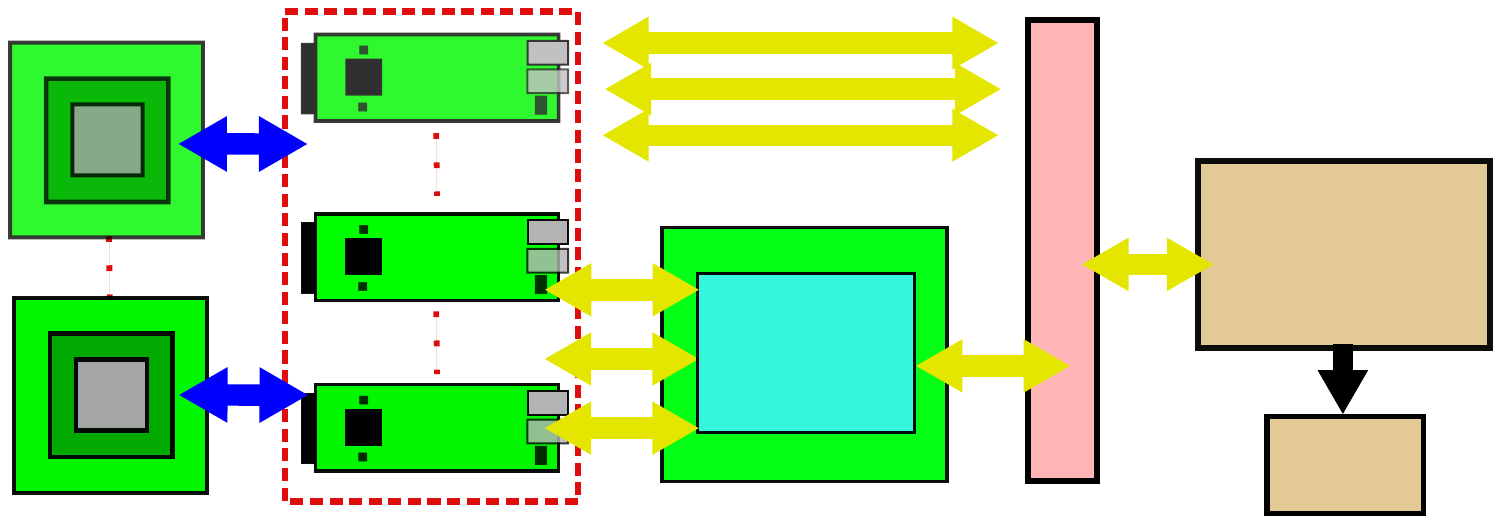}}
          \put(0,98){\color{black}\scriptsize{Sensors + FEE}}
          \put(56,103){\color{black}\scriptsize{DiRICH5d2}}
          \put(107,78){\color{black}\scriptsize\shortstack[c]{optical connection\\for data transfer\\and slow control}}
          \put(93,43){\color{black}\scriptsize{optical}}
          \put(121,35.3){\color{black}\scriptsize\shortstack[c]{\textbf{DCM}\\Dogma\\Control\\Module}}
          \put(152,43){\color{black}\scriptsize{optical}}
          \put(170,60){\begin{turn}{90}\color{black}\scriptsize{Switch}\end{turn}}
          \put(197,53){\color{black}\scriptsize\shortstack[c]{\textbf{DAQ PC}\\Slow control\\Event building\\Monitoring}}
          \put(208,22.5){\color{black}\scriptsize\shortstack[c]{Data\\storage}}
    %\shortstack[l]
    %      \put(181,132){\color{black}\footnotesize{pcCVD diamond}}
    %      \put(10,43){\begin{turn}{-8}\color{black}\footnotesize{Magnet}\end{turn}}
    %      \put(65,53){\begin{turn}{-8}\color{black}\footnotesize{Beam direction}\end{turn}}
    %      \put(69,25){\begin{turn}{-8}\color{black}\footnotesize{$\approx$ 195 cm}\end{turn}}
    %      \put(128,16){\begin{turn}{-8}\color{black}\footnotesize{1.3 cm}\end{turn}}
    %      \put(165,10){\begin{turn}{-8}\color{black}\footnotesize{8.7 cm}\end{turn}}
        \end{picture}
      \end{minipage}%
    \vspace{-0.5cm}
            \caption{DOGMA data acquisition system. The sensors, mounted on dedicated PCBs, are connected to the DiRICH5d2 boards, which feature FPGA-based leading-edge discriminators and TDCs. Communication with the DiRICH5d2 boards is fully optical and is controlled by the DOGMA Control Module (DCM). Data are transferred through a commercially available network switch to the DAQ computer for storage and further analysis.}
      \label{fig:dcm}
    \end{figure}

    The basic components of the DOGMA-based readout system used in this work are schematically shown in Fig.~\ref{fig:dcm}. The system consists of: (1) sensors mounted on PCBs equipped with two-stage signal amplification, (2) DiRICH5d2 front-end readout boards, (3) the DOGMA Control Module (DCM), (4) a network switch, and (5) a control and data acquisition computer. Event building and data handling are performed by the Data Acquisition Backbone Core (DABC) framework~\cite{joern2008,dabc}, with the load-sharing configuration controlled by DOGMA.

    %    \begin{figure}[b!]
    %        \centering
    %        \includegraphics[width=\columnwidth]{Images/dogma_general_sketch.pdf}
    %        \caption{Dogma data acquisition system. The sensors are connected to the DiRICH5d2 boards, which feature an %FPGA-based leading-edge discriminator and TDC. The readout of the DiRICH5d2 boards is purely optical.}
    %        \label{fig:dcm}
    %    \end{figure}
    \begin{figure}[t!]
        \centering
        \includegraphics[width=\columnwidth]{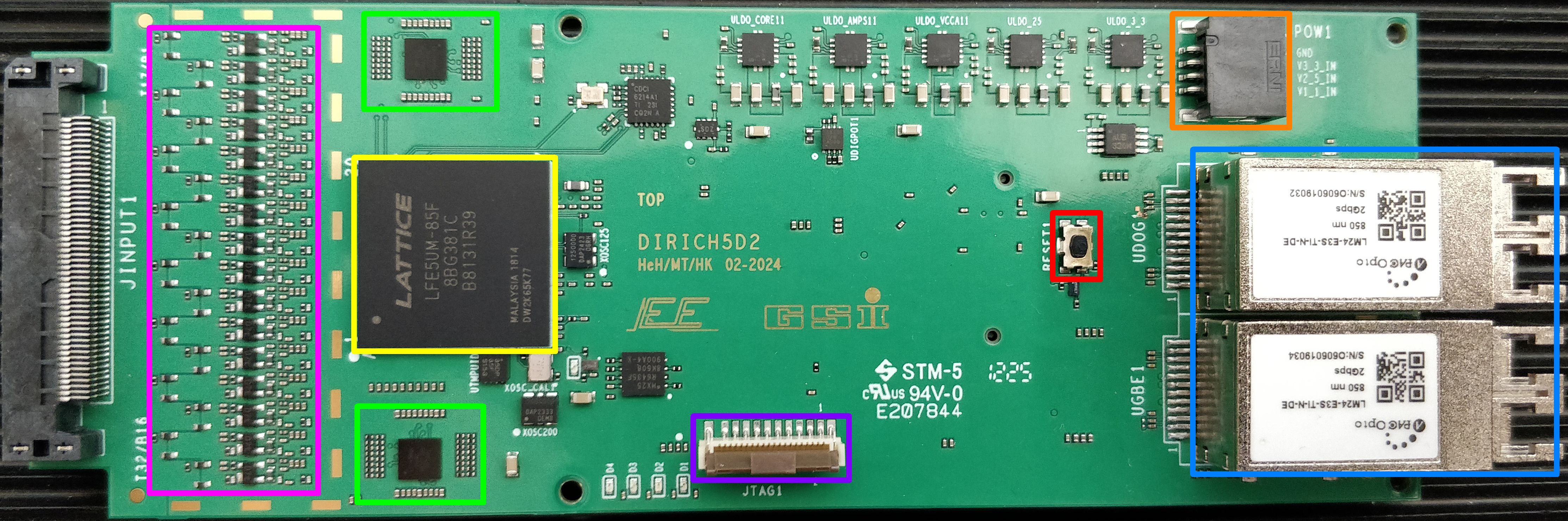}
        \caption{ Photograph of the DiRICH5d2 readout board (130 mm $\times$ 47 mm), with the main components highlighted: pre-amplification stage (pink), main FPGA (yellow), satellite FPGAs for threshold control (green), JTAG connector (violet), physical reset button (red), power connector (orange), and optical Small Form Factor (SFF) ports (blue).}
        \label{fig:dirich}
    \end{figure}

    \begin{figure}[t!]
        \centering
        \includegraphics[width=\columnwidth]{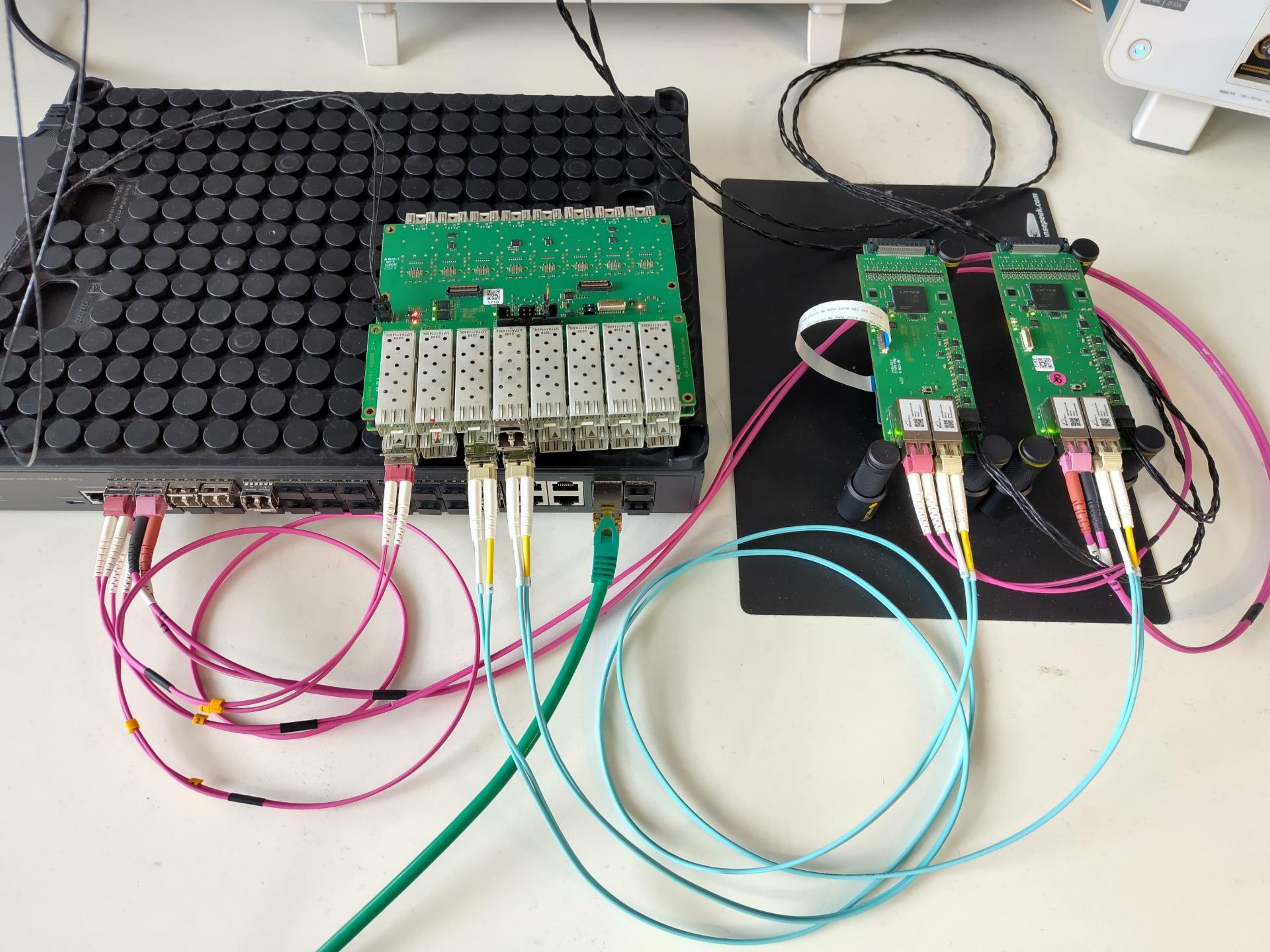}
        \caption{Photograph of a simple DOGMA system consisting of a DCM (center) and two DiRICH5d2 boards (right). The DOGMA TDC control path is indicated by the blue optical cables, while the data transmission path to the network switch is shown in pink.}
        \label{fig:dogma_con}
    \end{figure}

    The DCM serves as the master controller for the synchronous subsystem. It distributes a high-precision system clock (150~MHz with $<10$~ps RMS jitter), issues DAQ triggers and time-synchronization with deterministic latency and also receives trigger requests from end-points if they are configured to trigger the system depending on the hits they receive via an trigger algorithm implemented in the FPGAs.
    
    Analog signals are processed by the DiRICH5d2 board (Fig.~\ref{fig:dirich},~\ref{fig:dogma_con}), which features an integrated amplification stage with a variable gain of $G = 15-50$ (nominal is $G = 30$), followed by leading-edge discriminators and FPGA-implemented Time-to-Digital Converters (TDCs). The TDCs have a high timing precision for both leading edge and time-over-threshold measurements, with 8~ps and 14~ps RMS respectively \cite{ugur2015}. Discriminator thresholds are managed by two dedicated satellite FPGAs on the board, providing simultaneous readout for 32 channels. As the system relies on a leading-edge discriminator, a time-walk correction~\cite{felix2024} is required to achieve precise time measurements. For this purpose, the TDC records not only the Time-of-Arrival (ToA) but also the Time-over-Threshold (ToT). The ToT information enables an amplitude-dependent time-walk correction and additionally provides a valuable observable for detector performance monitoring, for example, for tracking changes in sensor response due to radiation damage. A single DCM can control up to 30 DiRICH5d2 boards. The system can be expanded by adding additional DOGMA-HUBs (same hardware as the DCM) connecting them via the DOGMA-link, providing a flexible scaling capability for larger detector systems. To minimize electromagnetic interference, signal pickup, and grounding issues, the system architecture relies completely on optical communication links to provide galvanic isolation. The DiRICH5d2 uses 1~Gb/s Ethernet. The new versions of the module (CERBERUS1/2) have 10~Gb/s links.
    
    The data acquisition side is also designed to be truly scalable. The achievable total data throughput is only limited by the bandwidth of the commercial Ethernet infrastructure used to transport the generated data. The controls system of DOGMA was designed to monitor (and change) intrinsic parameters of the running detector system, such as discriminator thresholds and channel count rates. Combined with a fast response time and a flexible discriminator threshold configuration, it enables a rapid and user-friendly threshold optimization procedure that can be completed within a few seconds. Direct access to the discriminator threshold settings and channel hit rates provides powerful online monitoring capabilities and enables efficient detector commissioning and operation.
    
    The TDC functionality is implemented in an FPGA and dedicated hardware calibration is required to achieve the desired timing performance. 
    The most important calibration step is the correction of the TDC integral non-linearity (INL), whose characteristics depend on silicon process variations, which differ between individual FPGAs, and on temperature. To perform this calibration, a dedicated on-board calibration pulse generation mechanism has been implemented and can be activated through a control register. In calibration mode, the DABC framework automatically identifies the calibration data, processes the acquired events, accumulates the required statistics, and generates the TDC calibration tables, which are subsequently applied during data taking. The fast threshold optimization procedure and the automated TDC calibration are key features of the DOGMA readout system. Together, they significantly simplify detector commissioning, reduce experiment preparation time, and ensure stable detector operation with high timing performance.
    
    In the CBM experiment, precise beam quality monitoring is essential. Therefore, an independent data path from the CBM T$_0$ detector is required. For this purpose, the measurement of channel-wise hit rates within well-defined time intervals is sufficient. The DOGMA system provides continuous access to count-rate information for each TDC channel with a maximum detectable input hit frequency of 150~MHz. This capability will be utilized for beam quality monitoring. Integration windows in the order of several microseconds are sufficient to provide accurate information about the beam conditions. The low-level controls data will be transferred via the DOGMA Link to a dedicated FPGA-based monitoring board, where the beam quality evaluation and beam-abort logic of the CBM experiment will be implemented.

\section{Mechanical Integration and Alignment Concept of the CBM T$_0$ Detector}
    \label{mechanics}
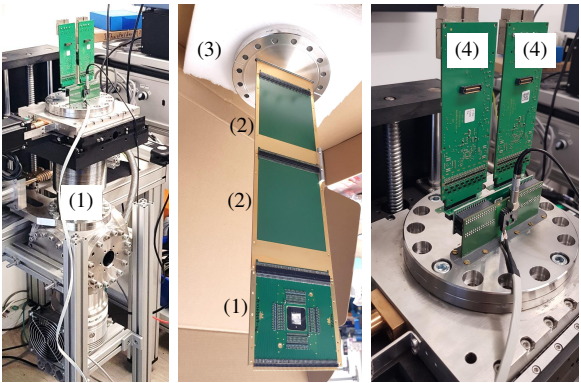
\begin{figure}[htb]
      \centering
      \setlength{\unitlength}{1mm} % Sets picture units to millimeters
      \begin{picture}(100, 55)(0, 0) % (Width, Height) in mm
        % Insert cropped/scaled PDF
        \put(0, 0){\includegraphics[trim=0mm 0mm 0mm 0mm, clip, width=\linewidth]{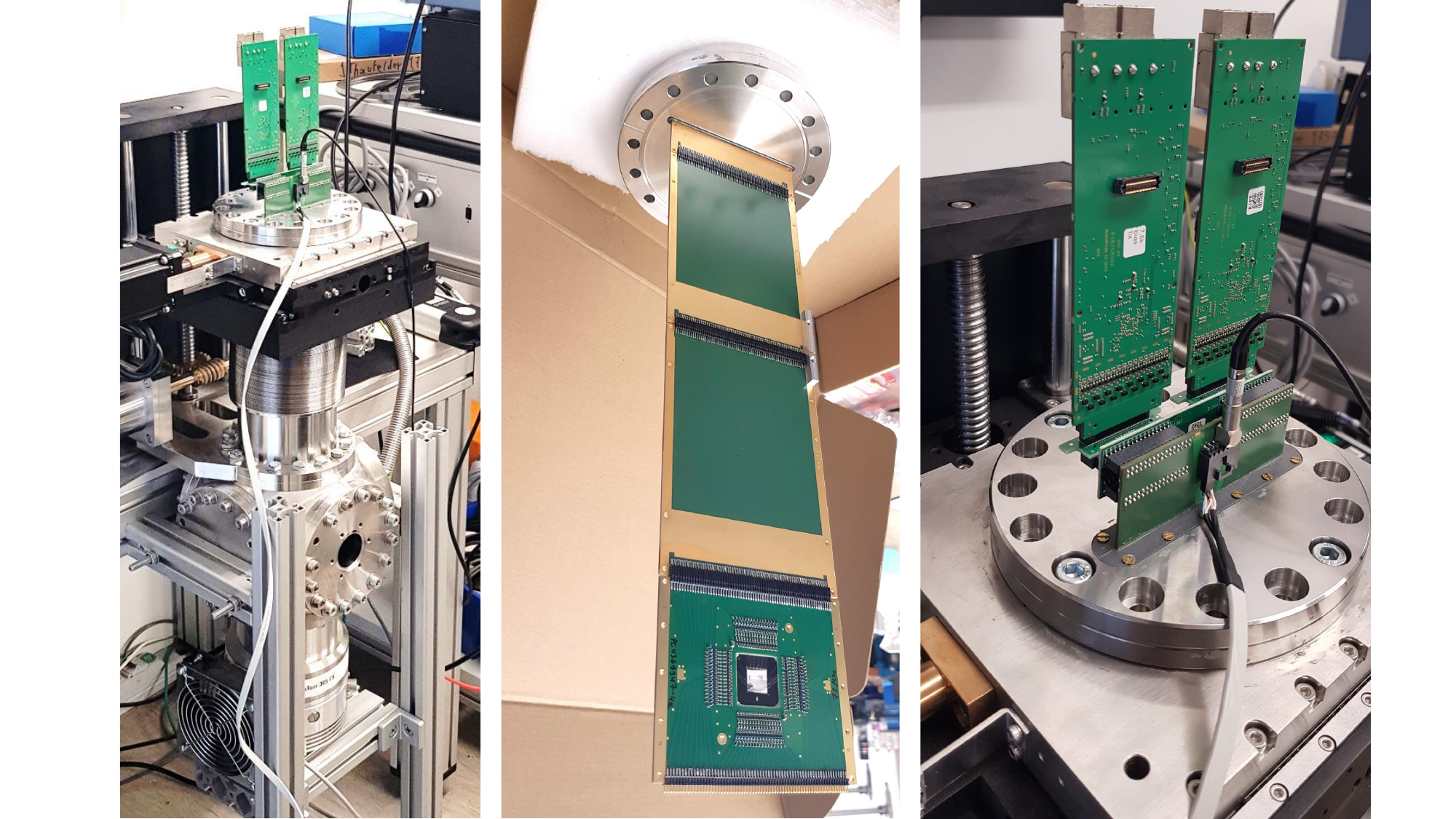}}
        
        % Example overlay text using millimeter coordinates
        %\put(18, 23){\color{black}\footnotesize{(1)}}
        \put(15, 23){\colorbox{white}{\footnotesize{(1)}}}
        \put(36.5, 9){\color{black}\footnotesize{(1)}}
        \put(37, 23){\color{black}\footnotesize{(2)}}
        \put(37.3, 33){\color{black}\footnotesize{(2)}}
        \put(33, 43){\color{black}\footnotesize{(3)}}
        \put(66, 43){\colorbox{white}{\footnotesize{(4)}}}
        \put(75, 43){\colorbox{white}{\footnotesize{(4)}}}

      \end{picture}
      
      \caption{Hardware integration of the CBM $\mathrm{T}_0$ beam-test setup. (Left) Vacuum manipulator mounted on the test stand (1), housing the $\mathrm{T}_0$ sensor assembly. (Middle) Close-up of the sensor package, showing the diamond detector mounted on its front-end preamplifier PCB (1) and connected via two interface boards (2) to the internal side of the vacuum feedthrough on a standard DN100 (100 mm) flange (3) . (Right) Atmospheric side of the vacuum flange, with two DiRICH5d2 readout boards (4) directly connected to the external feedthrough connectors.}
      \label{fig:vacuum_installation}
    \end{figure}

    To ensure reliable operation during CBM physics runs, the CBM T$_0$ detector must be integrated directly into the beam line vacuum system. The sensor assembly will be housed inside a custom beam pipe section and mounted on a high-precision 3D vacuum manipulator. The selected vacuum manipulator provides a vacuum tube with an inner diameter of 10 cm and a distance of approximately 30–40 cm between the mounting flange and the beam axis. The CBM T$_0$ sensor must be positioned precisely on the beam axis. To achieve this, the sensor will be mounted on a dedicated holding PCB, which is mechanically supported by an extension PCBs connecting the holding PCB to the vacuum feedthrough PCB. The mechanical arrangement of the sensor support is illustrated in Fig.~\ref{fig:vacuum_installation}-Center. The vacuum feedthrough PCB is inserted into a slot milled into the vacuum flange and permanently sealed using a high-vacuum epoxy adhesive to ensure vacuum tightness. In addition to providing vacuum tightness and mechanical stability, the feedthrough PCB routes all electrical connections between the detector inside the vacuum chamber and the front-end electronics outside the vacuum. These connections include the analog signal lines and the low- and high-voltage supply lines. The overall distance from the flange surface to the center of the sensor is approximately 35 cm. This geometry places the sensor at the nominal detector position when the vacuum manipulator is operated at the center of its travel range, thereby providing sufficient adjustment margin in both directions for detector alignment.  Outside the vacuum vessel, the analog detector signals are connected directly to the DiRICH front-end readout boards, where signal amplification, discrimination, and digitization are performed (Fig.~\ref{fig:vacuum_installation}-Right). For the T$_0$ detector equipped with a $6 \times 6$ pad-metallized diamond sensor, the 36 readout channels are processed by two DiRICH boards, each serving 18 detector channels.
    
    In the CBM experiment, precise transverse alignment is particularly critical when operating in magnetic fields. The target is fixed at a position of 44 cm upstream of the magnet's central point, where the magnetic field strength reaches 77\% of its maximum value. At the location of the T$_0$ detector, 1 m upstream of the target, the fringe field from the CBM dipole magnet is still present at a level of approximately 5\% of the maximum field strength. Depending on the ion species, beam energy, and magnetic field strength, the magnetic field induces a horizontal beam deflection of several millimeters at the CBM target position. To ensure the beam hits the center of the target, the incident beam trajectory has to be steered horizontally prior to reaching the target region. To track this trajectory shift and maintain optimal overlap with the beam spot, particularly during light-ion campaigns, the manipulator will be mounted horizontally, enabling motorized real-time position adjustments of the T$_0$ sensor.
    
\section{Performance Evaluation of the $\mathrm{T}_0$ System with Pad Metallization using Light Ions.}
    \label{perf_evaluation}
    To validate the operational performance of the CBM T$_0$ detector system with pad metallization, beam tests were performed using several light-ion species over a broad range of beam energies. The diamond T$_0$ detector used in this study was a sensor previously operated in mCBM test runs at GSI Darmstadt with high-intensity, well-focused $^{107}$Ag and $^{197}$Au ion beams. During these measurements, the sensor was exposed to significant irradiation, resulting in a noticeable reduction of its detection efficiency. At that time, the mCBM setup did not include additional amplification stages in the readout chain.
    The previously irradiated sensor was selected for the present study to evaluate the performance of the CBM T$_0$ readout system under realistic degraded detector conditions. The objective was to demonstrate that the implementation of additional amplification stages in the readout electronics can compensate for the reduced signal amplitude and extend the operational lifetime of diamond detectors equipped with pad metallization. This approach follows previous observations obtained with strip metallization, where the benefits of additional amplification stages were demonstrated in Ref.~\cite{pie2025}.
    This chapter describes the experimental setup and the offline calibration procedure, including channel-to-channel time offset corrections, ToT calibration, and time-walk corrections. Finally, the detection efficiency and timing performance obtained for the different ion species and beam energies are presented and discussed.
    
    \begin{figure}[htb]
      \centering
      \hspace*{0mm}
      \begin{minipage}[t]{0.50\textwidth}
        \centering
        \hspace*{0mm} %% shift all to the right
        %\begin{picture}(0.55\textwidth,185)(0,0)
        \begin{picture}(\columnwidth,155)(0,0)
          \put(0,0){\includegraphics[width=\columnwidth, trim= 0  0 0 0, clip ]{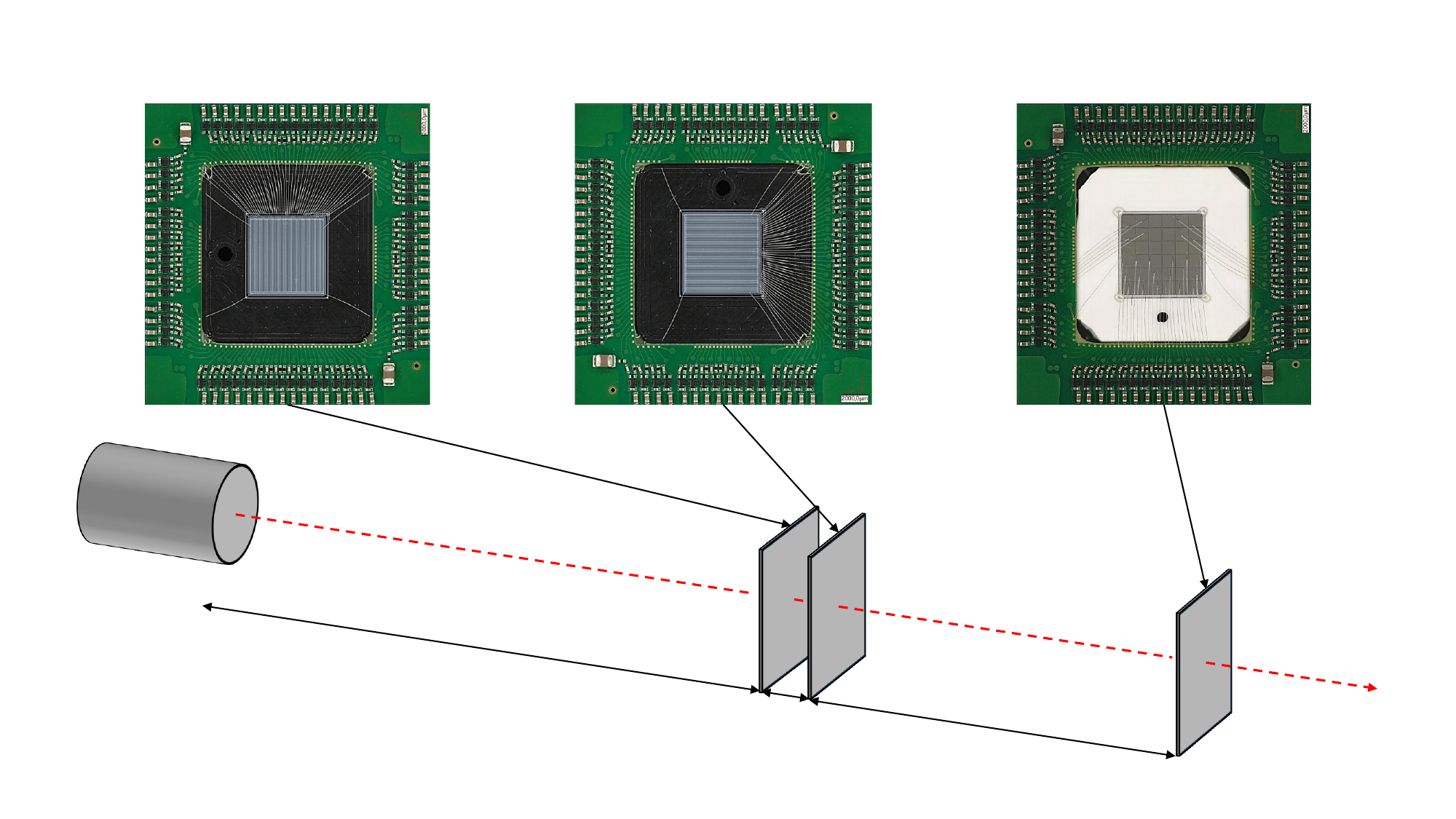}}
          \put(36,132){\color{black}\footnotesize{LGAD X}}
          \put(115,132){\color{black}\footnotesize{LGAD Y}}
          \put(181,132){\color{black}\footnotesize{pcCVD diamond}}
          \put(10,43){\begin{turn}{-8}\color{black}\footnotesize{Magnet}\end{turn}}
          \put(65,53){\begin{turn}{-8}\color{black}\footnotesize{Beam direction}\end{turn}}
          \put(69,25){\begin{turn}{-8}\color{black}\footnotesize{$\approx$ 195 cm}\end{turn}}
          \put(128,16){\begin{turn}{-8}\color{black}\footnotesize{1.3 cm}\end{turn}}
          \put(165,10){\begin{turn}{-8}\color{black}\footnotesize{8.7 cm}\end{turn}}
        \end{picture}
      \end{minipage}%
            \caption{Sketch of LGAD and diamond telescope setup, the $1\times1$~cm$^2$ sensors are oriented as seen by the beam.}
      \label{fig:He_C_test_setup}
    \end{figure}
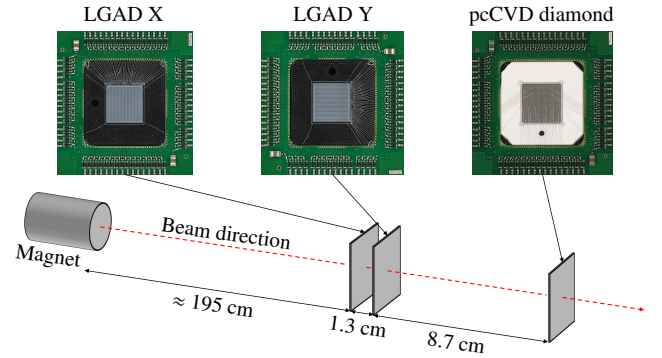

    \subsection{Experimental Setup and Beam Conditions}
        The beam test was performed at EBG MedAustron (Wiener Neustadt, Austria) using $^4$He beams with kinetic energies of 73.3, 124.4, 147.7, and 177.7~$A$MeV, and a $^{12}$C beam with an energy of 402~$A$MeV. The CBM T$_0$ diamond detector with pad metallization was installed downstream of the ion Computed Tomography (iCT) setup described in Refs.~\cite{felix2023,felix2024}. The ionCT setup consists of 12 LGAD detector planes providing precise tracking and timing information. For the present performance study, the last two LGAD planes were used as reference detectors, forming a beam telescope together with the diamond detector, as illustrated schematically in Fig.~\ref{fig:He_C_test_setup}. During the test experiment, signals from sensor 11 of the ionCT setup, which is the first sensor of our telescope, were used as the trigger for the DOGMA DAQ system.
        
        The reference detectors were LGAD sensors produced by Fondazione Bruno Kessler (FBK), featuring an active area of 8.6~$\times$~8.6~mm$^2$ and a total thickness of 200~\textmu m. Similar sensors have previously been employed for electron beam diagnostics~\cite{vadym2022} and as T$_0$ detectors for proton beam experiments in HADES~\cite{willy2022}. Each LGAD sensor comprises 45 readout strips with a pitch of 192~\textmu m. Arranging the strip orientations of the two sensors perpendicular to each other provides an effective segmentation of 2025 virtual pixels, enabling precise measurements of both the particle impact position and timing. All sensors were glued and wire-bonded to dedicated front-end PCBs equipped with AC coupling and a two-stage amplification circuit~\cite{pie2025}. The PCBs were designed to accommodate sensors with active areas of up to 2$\times$2~cm$^2$. Since the sensors used in this study have dimensions of only 1$\times$1~cm$^2$, custom 3D-printed Poly-Lactic Acid (PLA) holders were used to position the sensors accurately on the PCBs. The holders are shown in Fig.~\ref{fig:He_C_test_setup} (black for the LGAD sensors and white for the diamond detector). The amplified signals from the front-end PCBs were transmitted to DiRICH5d2 readout boards, where they underwent further amplification, discrimination, and time digitization by the integrated TDC. To accommodate the full channel count, two DiRICH5d2 boards were used for each detector, allowing the readout of all 45 LGAD channels or all 36 channels of the diamond detector, respectively.
    
    \subsection{Calibration and Data Analysis Procedures}
        \label{calib}
        The raw timestamps recorded by the readout electronics cannot be used directly for the detector performance evaluation. Several detector- and electronics-related effects influence the measured hit times and therefore require dedicated offline corrections and event reconstruction procedures. The most important effects are discussed below.
        
        First, the front-end electronics employ a leading-edge discriminator to determine the signal arrival time. Since the threshold crossing depends on both the signal arrival time and its amplitude, pulses with different amplitudes cross the fixed discriminator threshold at different times, resulting in the well-known time-walk effect. The signal amplitude information is encoded in the ToT measurement, allowing this effect to be corrected offline by applying an appropriate ToT-dependent time correction. Second, the individual readout channels differ in their electrical signal path lengths, including contributions from the detector routing, PCB traces, cables, and readout electronics. These differences introduce channel-dependent timing offsets that must be determined during calibration and removed to establish a common time reference for all channels. Finally, the combination of segmented strip and pad detectors, multi-stage amplification, and highly integrated readout electronics leads to correlated responses among neighbouring channels. These correlations arise from several mechanisms, including charge sharing in the detector, capacitive coupling between adjacent electrodes, and electronic crosstalk within the amplification and discrimination stages. Consequently, a single particle may produce multiple spatially and temporally correlated hits that form a cluster rather than an isolated hit. The identification and characterization of such clusters constitute an essential part of the analysis. Appropriate clustering algorithms are used to identify the hit corresponding to the traversing particle while rejecting parasitic signals originating from coupling and crosstalk.
        
        The time-walk and channel-to-channel time offset corrections are performed in a single calibration procedure after applying a preliminary clustering algorithm. The strip detector geometry with perpendicular strip orientations is particularly advantageous for this calibration, as a single strip of one detector can be used to calibrate all strips of the second detector, and vice versa. The identification of clustered hits is especially important for the time-walk correction. Signals originating from charge sharing, capacitive coupling or electronic cross-talk are typically smaller in amplitude than the primary particle signal. Consequently, they exhibit a larger delay in the threshold crossing time due to the time-walk effect. If not properly identified and rejected, such signals introduce additional broadening of the time-difference distributions and deteriorate the extracted timing performance. At the same time, the clustering information provides an efficient method for distinguishing genuine particle hits from noise and parasitic signals, thereby improving the purity of the selected hit sample. The cluster-finding algorithm developed for this analysis identifies hits that are correlated in time and, where applicable, in position within a single event and groups them into clusters. The hit with the largest ToT value within a cluster is considered the primary particle hit and defines the cluster position; this hit is referred to as the local maximum.
        
\begin{figure}[t!]
            \centering
            \begin{subfigure}[c]{0.45\columnwidth}
            \centering
            \caption{\label{fig:c_totvstdiff_raw} 402~$A$MeV $^{12}$C before corrections}
            \vspace{-0.0cm} 
            \includegraphics[width=\linewidth, trim=0 0 0 1.0cm, clip]{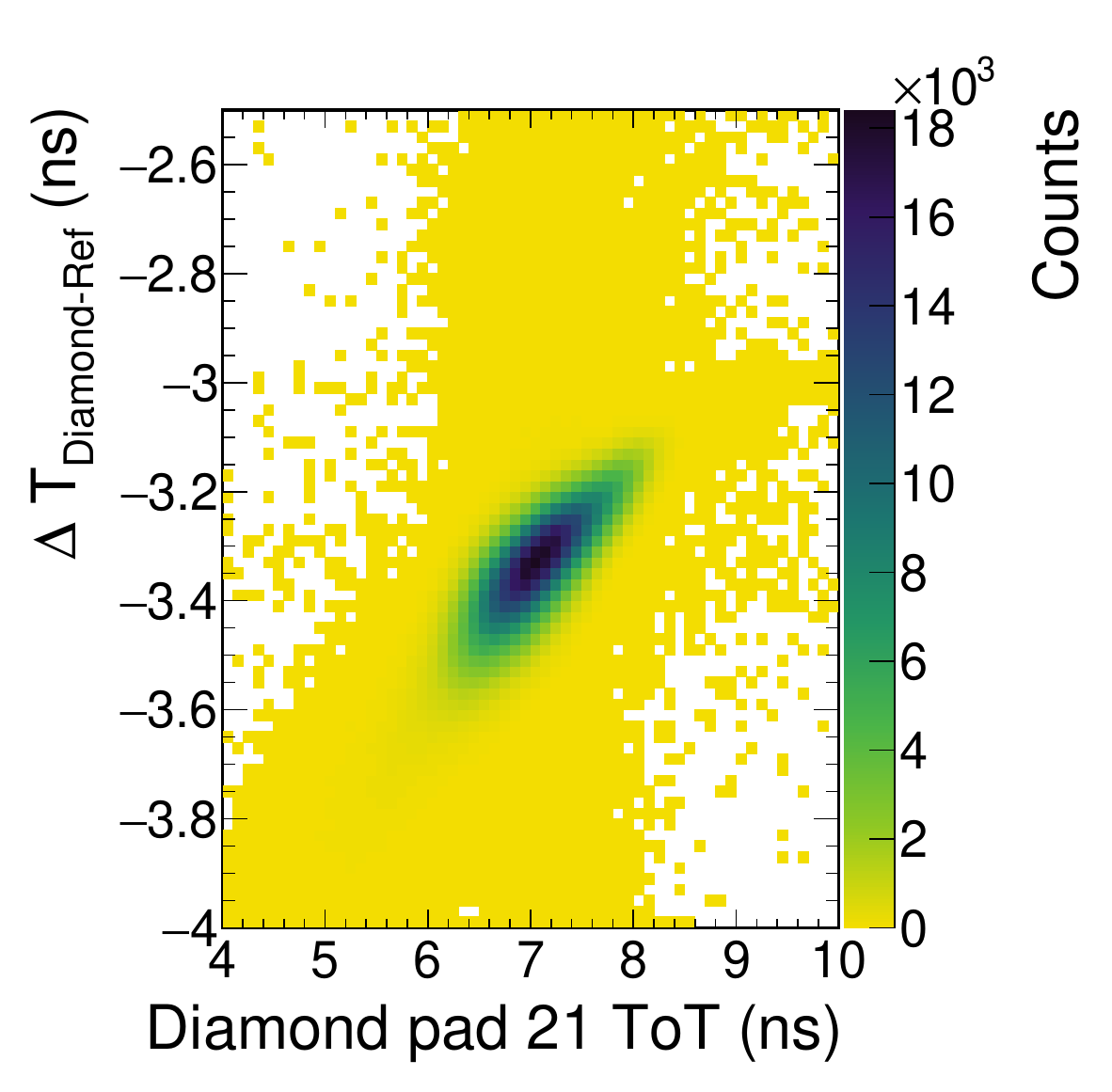}
            \end{subfigure}
            %\hfill
            \begin{subfigure}[c]{0.45\columnwidth}
            \centering
            \caption{\label{fig:c_totvstdiff_corr} 402~$A$MeV $^{12}$C after corrections}
            \vspace{0cm}
            \includegraphics[width=\linewidth, trim=0 0 0 1cm, clip]{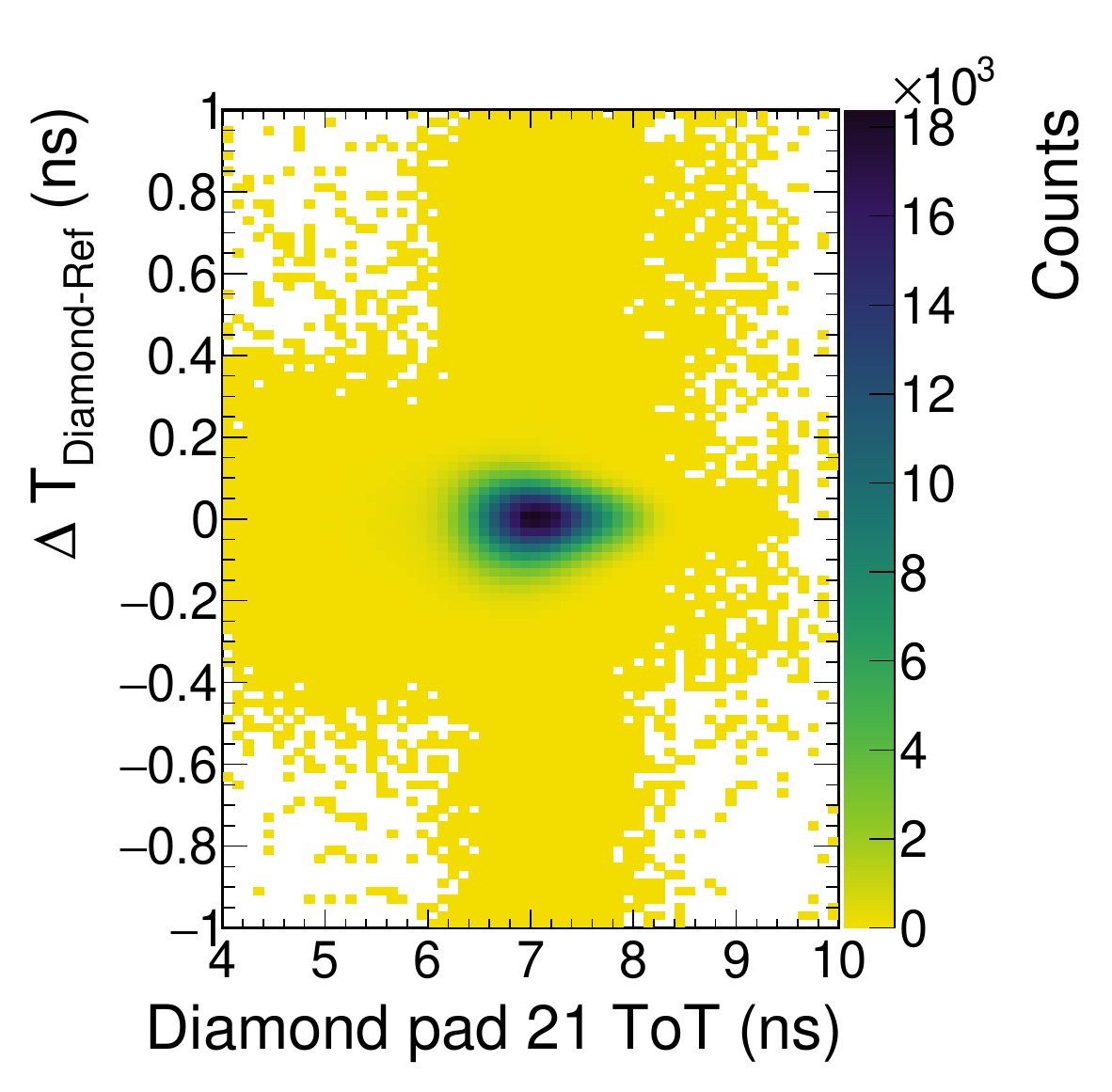}
            \end{subfigure}
            \vfill
            \begin{subfigure}[c]{0.45\columnwidth}
            \centering
            \caption{\label{fig:c_tdiff_raw} 402~$A$MeV $^{12}$C before corrections}
            \includegraphics[width=\linewidth, trim=0 0 0 1cm, clip]{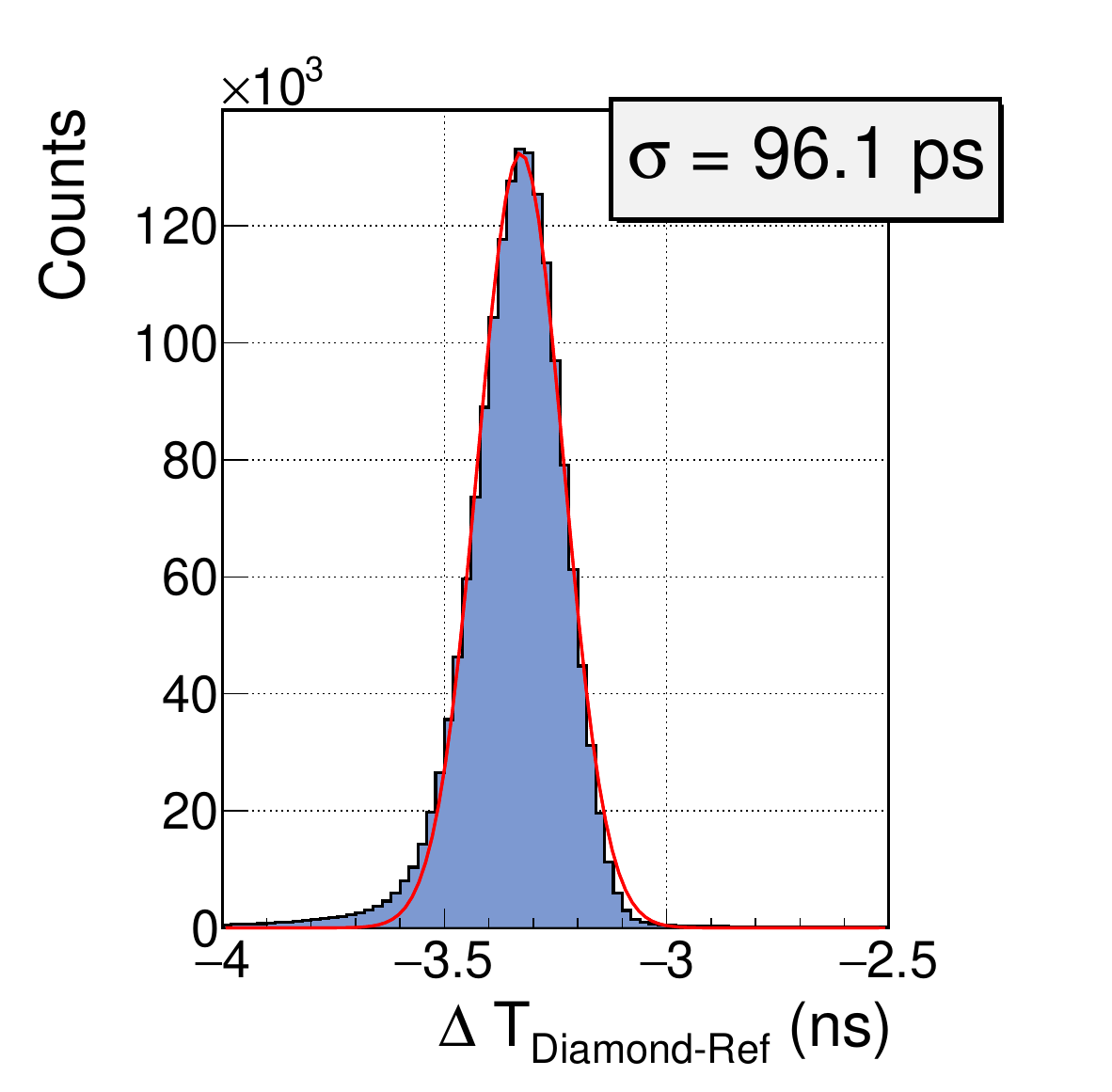}
            \end{subfigure}
            %\hfill
            \begin{subfigure}[c]{0.45\columnwidth}
            \centering
            \caption{\label{fig:c_tdiff_corr} 402~$A$MeV $^{12}$C after corrections}
            \includegraphics[width=\linewidth, trim=0 0 0 1cm, clip]{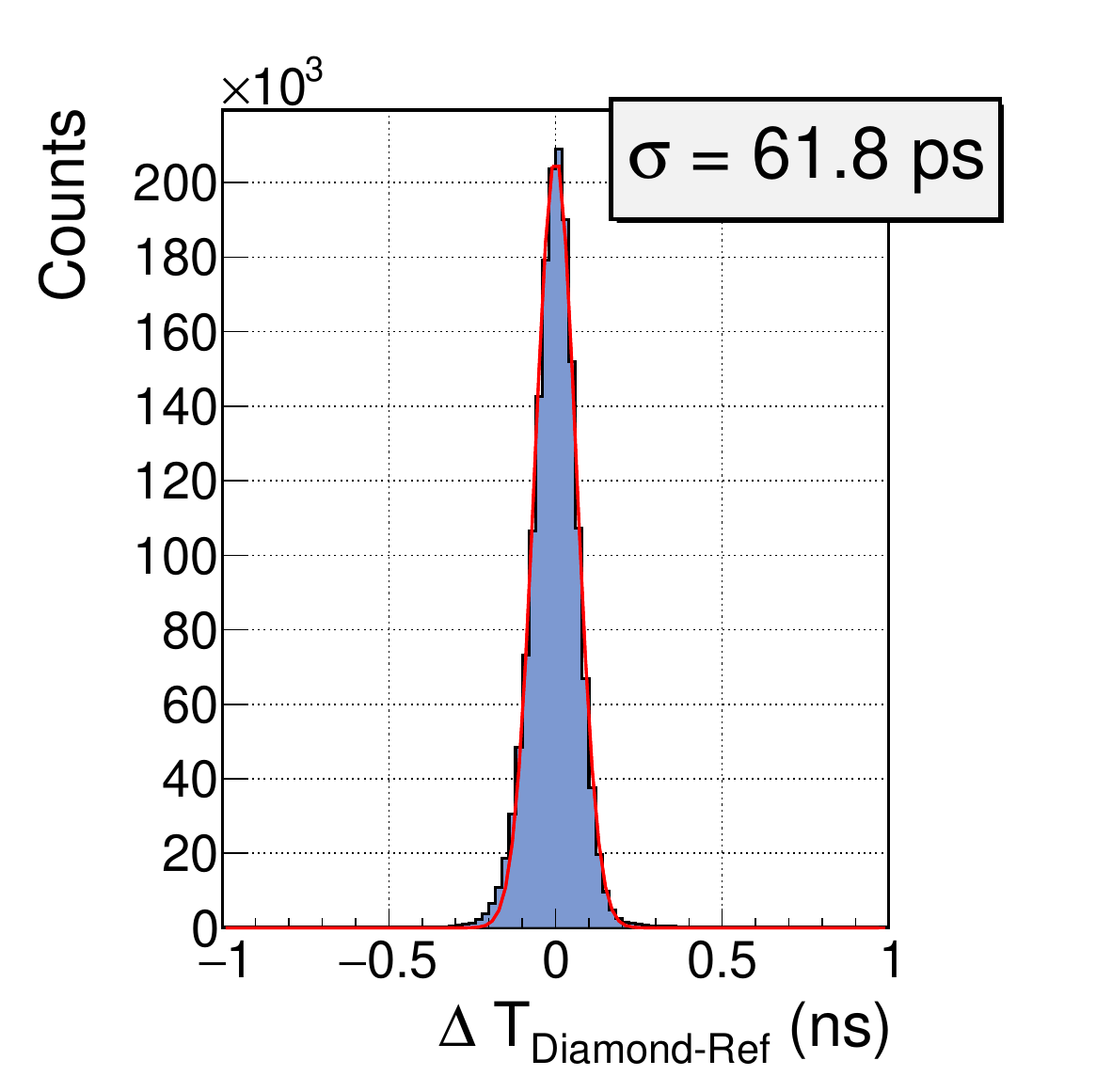}
            \end{subfigure}
            %\vfill
            \begin{subfigure}[c]{0.45\columnwidth}
            \centering
            \caption{\label{fig:he_sigma} 73~$A$MeV $^{4}$He}
            \includegraphics[width=\linewidth, trim=0 0 0 1.1cm, clip]{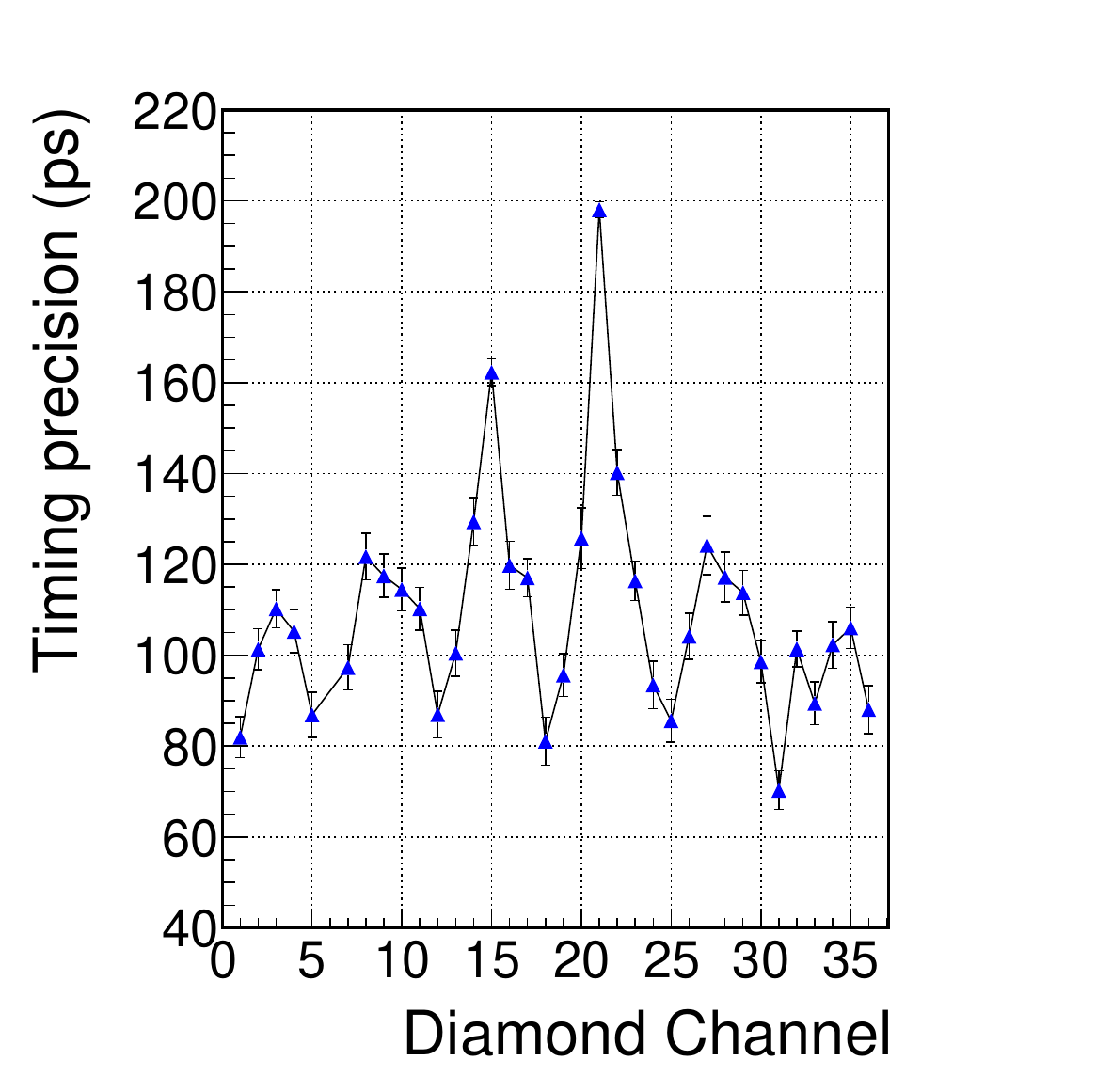}
            \end{subfigure}
            %\hfill
            \begin{subfigure}[c]{0.45\columnwidth}
            \centering
            \caption{\label{fig:c_sigma} 402~$A$MeV $^{12}$C}
            \includegraphics[width=\linewidth, trim= 0 0 0 1.1cm, clip]{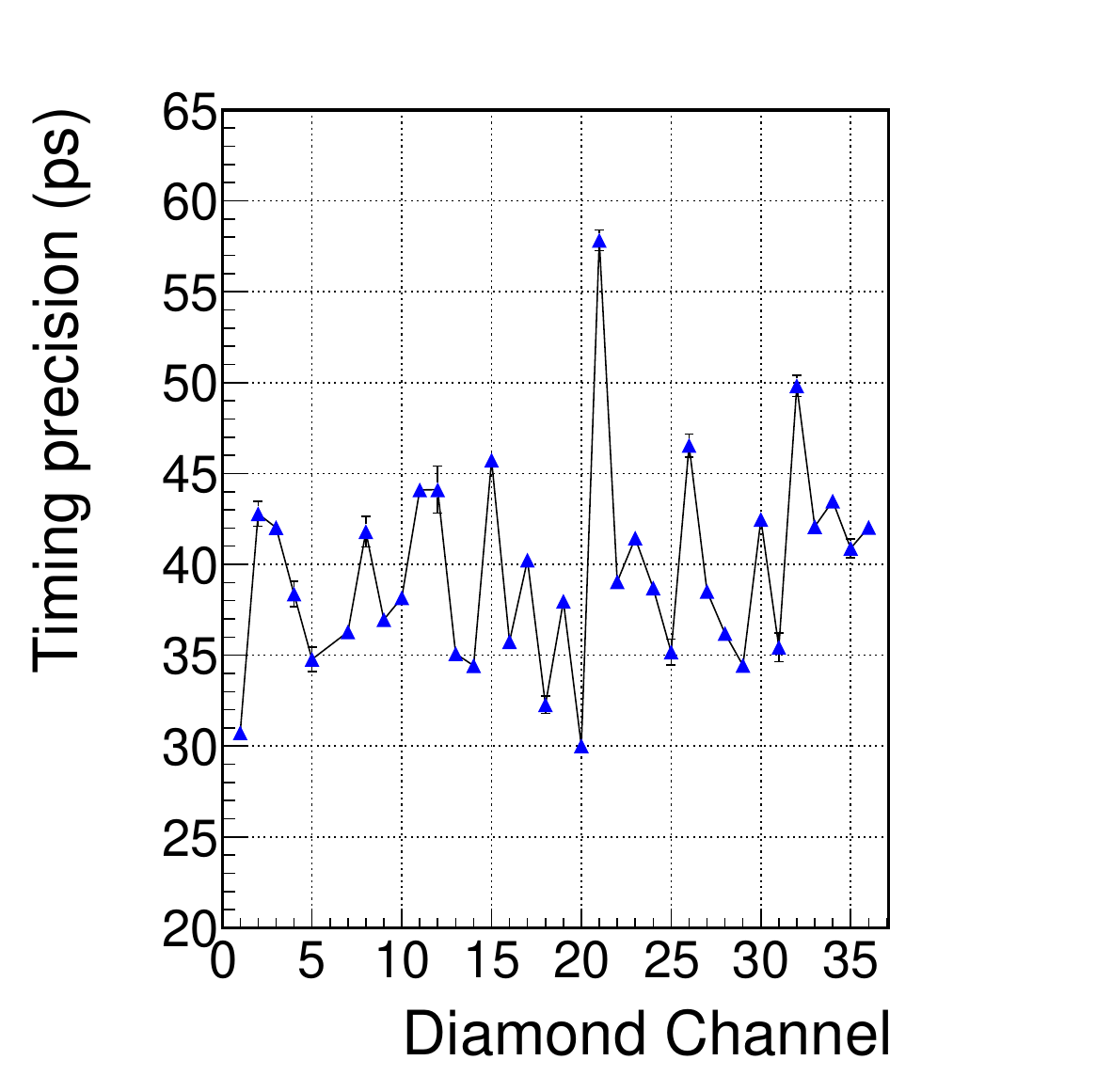}
            \end{subfigure}
            \caption{Top: ToT dependent distributions of $\Delta T_{\mathrm{Diamond\text{-}Ref}}$ for the most damaged diamond pad. Note that the colors are inverted compared to the efficiency maps.\\
            Middle: $\Delta T_{\mathrm{Diamond\text{-}Ref}}$ projections over the shown ToT range with Gaussian fit, showing $\sigma_{\mathrm{Total}}$.\\
            Bottom: $\sigma_{\mathrm{Diamond}}$ distributions for all diamond channels, calculated using Eq. (\ref{sigma_diam}).}
            \label{fig:timing}
        \end{figure}

        After the preliminary clustering step, the time difference between a selected channel of the first detector and each channel of the second detector is evaluated as a function of the ToT value. These two-dimensional distributions are analyzed separately for each channel to determine the ToT-dependent time-walk correction parameters and the corresponding channel timing offsets. An example of the time-difference distribution before and after applying the correction is shown in Fig.~\ref{fig:timing}-Top.
        
    \subsection{Diamond Detection Efficiency Evaluation}
        \label{results}
        For the estimation of the $\mathrm{T}_0$ sensor efficiency, two LGAD sensors with perpendicular strip orientations were used as reference detectors. The analysis procedure closely follows the approach described in~\cite{pie2025}. After applying the complete offline correction chain, including time calibration, time-walk correction, and local-maximum selection, the beam particle positions (X/Y) were reconstructed based on coincident signals (within a $\pm10$~ns time window) detected in both reference LGAD stations. Subsequently, for each reconstructed beam particle in the X/Y LGAD system, the presence of a signal in the CBM $\mathrm{T}_0$ diamond detector was checked using the same $\pm10$~ns time coincidence window. The diamond detector efficiency was determined by calculating the ratio between the X/Y hit map obtained from events where signals were detected in all three sensors and the hit map obtained from events where only the reference LGAD sensors were required.
        
\begin{figure}[b!]
            \centering
            \begin{subfigure}[c]{0.45\columnwidth}
            \centering
            \caption{\label{fig:he_177_eff} 177~$A$MeV He}
            \includegraphics[width=\linewidth, trim=0 0 0 1cm, clip]{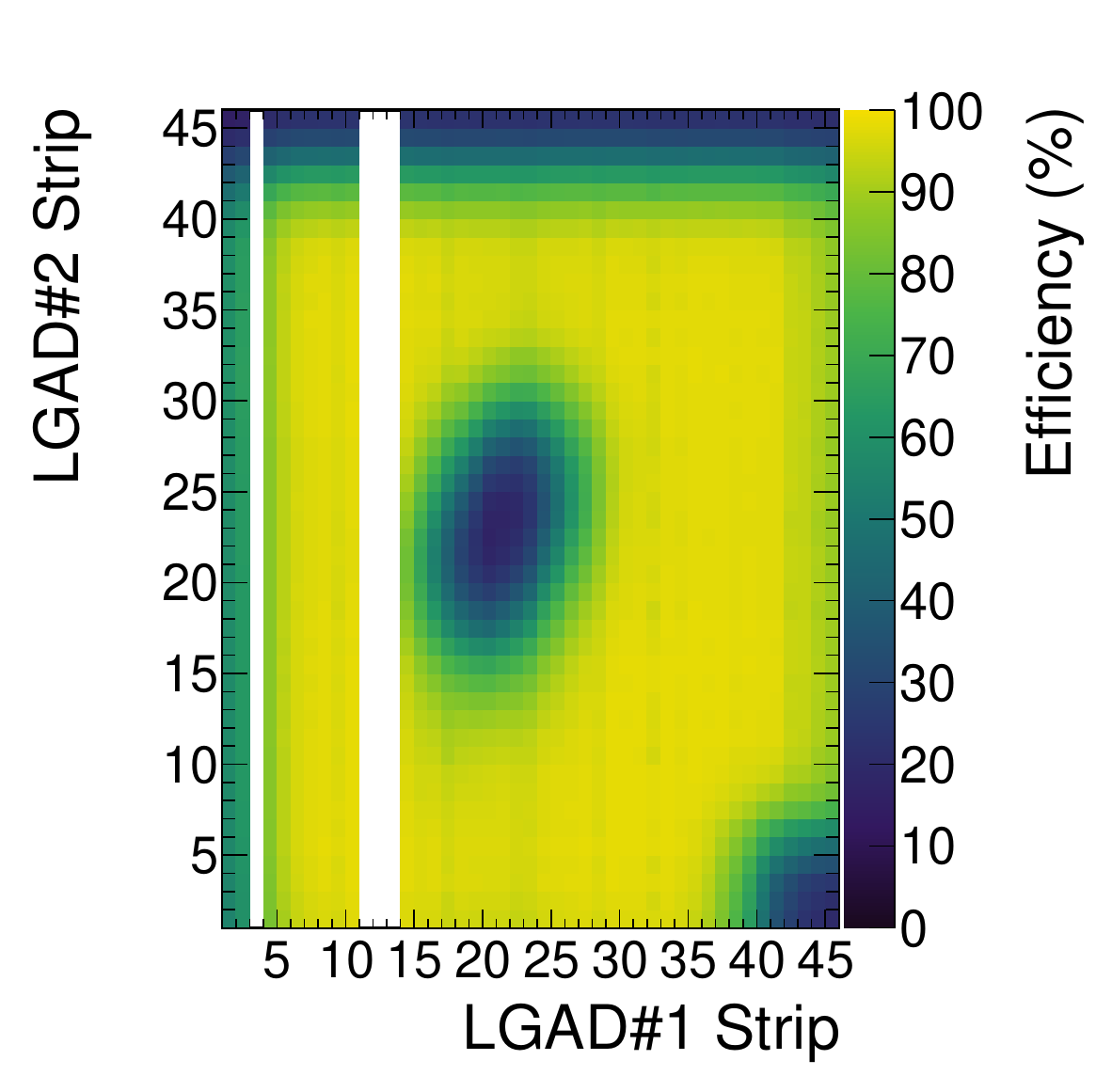}
            \end{subfigure}
            %\hfill
            \begin{subfigure}[c]{0.45\columnwidth}
            \centering
            \caption{\label{fig:he_124_eff} 124~$A$MeV He}
            \includegraphics[width=\linewidth, trim=0 0 0 1cm, clip]{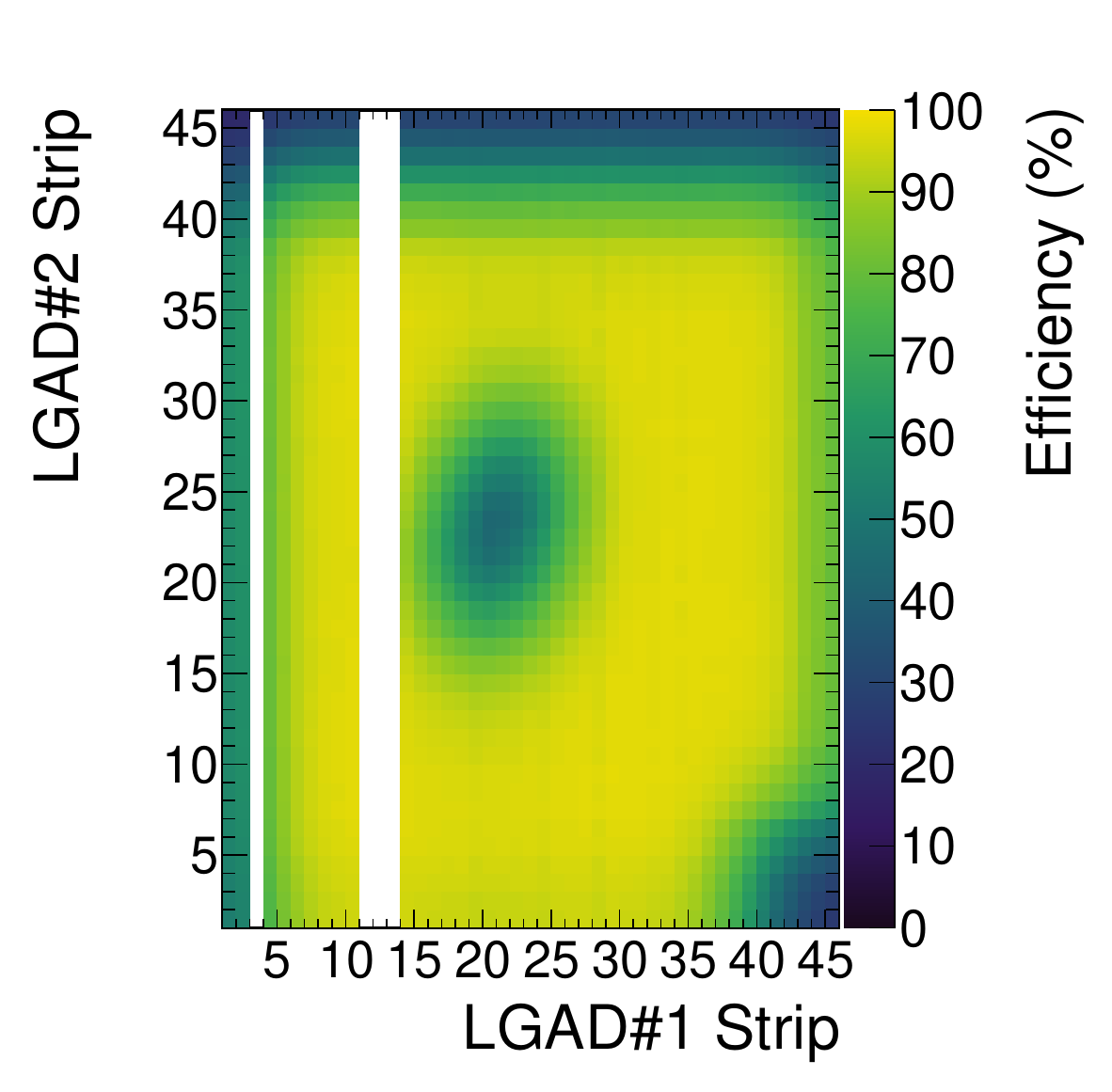}
            \end{subfigure}
            \vfill
            \begin{subfigure}[c]{0.45\columnwidth}
            \centering
            \caption{\label{fig:he_73_eff} 73~$A$MeV He}
            \includegraphics[width=\linewidth, trim=0 0 0 1cm, clip]{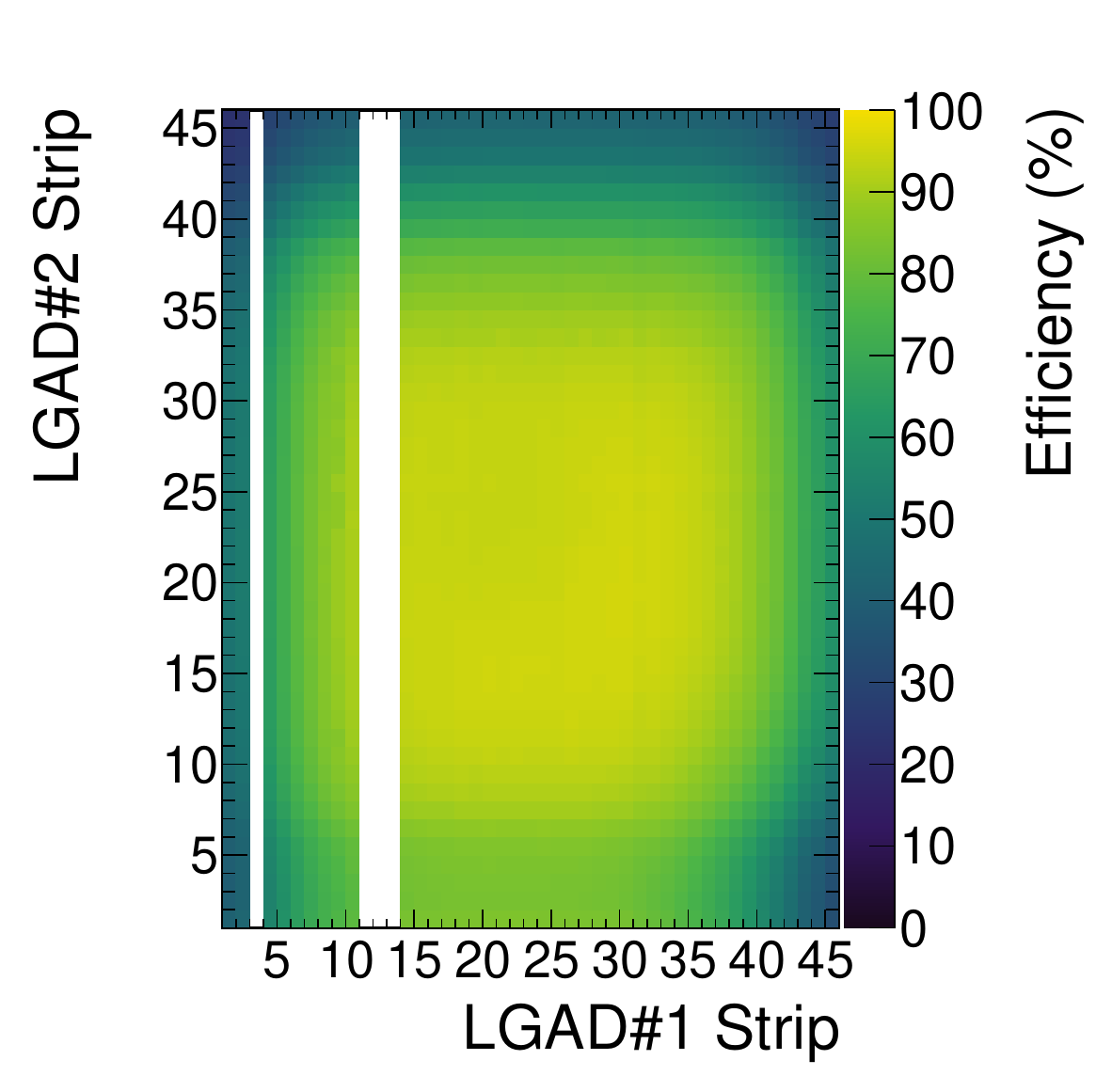}
            \end{subfigure}
            %\hfill
            \begin{subfigure}[c]{0.45\columnwidth}
            \centering
            \caption{\label{fig:c_402_eff} 402~$A$MeV C}
            \includegraphics[width=\linewidth, trim=0 0 0 1cm, clip]{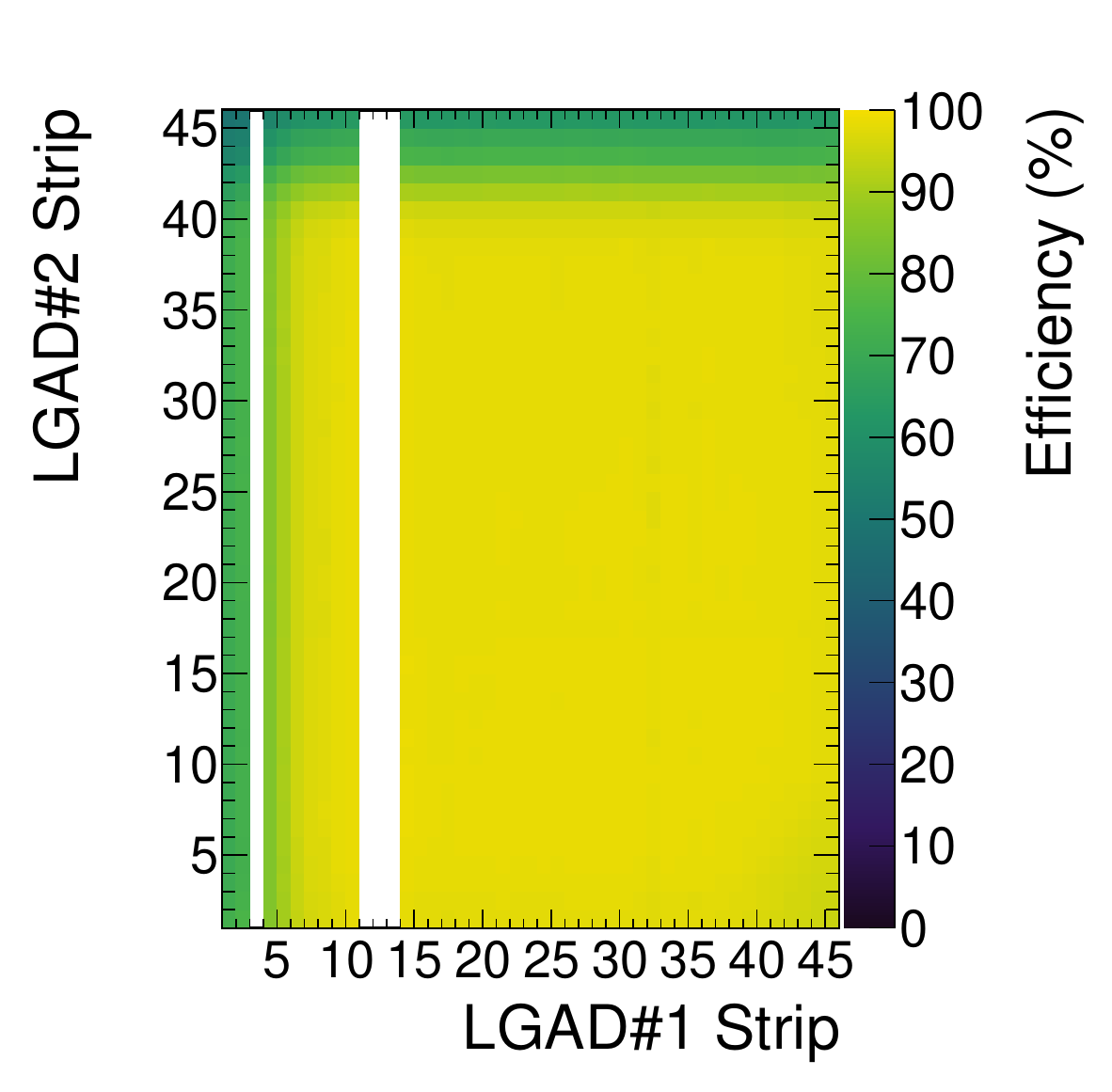}
            \end{subfigure}
            \caption{Efficiency maps for different $^4$He beam energies (a-c) and a $^{12}$C beam (d), ordered by increasing $dE/dx$.}
            \label{fig:eff_maps}
        \end{figure}

        The results of these analysis steps for selected $^4$He beam energies of 177, 124, and 73~$A$MeV, as well as for 402~$A$MeV $^{12}$C ions, are shown in Fig.~\ref{fig:eff_maps}. It can be observed that at the highest $^4$He energies of 177~$A$MeV (Fig.~\ref{fig:he_177_eff}) and 124~$A$MeV (Fig.~\ref{fig:he_124_eff}), a clear efficiency drop is visible, corresponding to the region affected by radiation damage. However, the inefficiency in the center of the damaged area gradually decreases with decreasing particle energy and, consequently, increasing energy loss per unit length ($dE/dx$), until it almost completely disappears for 73~$A$MeV $^4$He ions (Fig.~\ref{fig:he_73_eff}). For the $^{12}$C beam (Fig.~\ref{fig:c_402_eff}), no reduction in detection efficiency is observed in the irradiated region of the sensor.
    
    \subsection{Diamond Detector Time Precision}
        \label{timing}
        Using the reference detectors in our setup (Fig.~\ref{fig:He_C_test_setup}), we could precisely investigate the time precision of the CBM T$_0$ sensor for two beam conditions, $^4$He at 73~$A$MeV and $^{12}$C at 402~$A$MeV, for which the detector efficiency was measured to be close to 100\%. After applying the calibration procedures described in Chapter~\ref{calib} to the data from all three sensors in the setup, the time precision determination was performed.
        In this procedure, knowing that all pads exhibit an efficiency close to 100\%, the time precision estimation can be performed separately for each pad by requiring that the local maximum hit is located within the selected diamond pad. The corresponding diamond ToA differences with respect to the average ToA of the reference LGAD sensors are then calculated and plotted according to Eq.~\ref{delta_tdiff}.
        
        To extract the intrinsic timing precision of the diamond sensor ($\sigma_{\mathrm{Diamond}}$) without assuming identical performance for the reference detectors, a two-step variance evaluation was performed using the two LGAD sensors (LGAD$_1$ and LGAD$_2$).
        First, the relative timing difference between the two reference detectors was measured:
        
        \begin{equation}
            \Delta T_{\mathrm{ref}} = \mathrm{ToA}_{\mathrm{LGAD1}} - \mathrm{ToA}_{\mathrm{LGAD2}}
            \label{delta_T}
        \end{equation}
        The variance of this distribution combines the individual, unconstrained reference precisions ($\sigma_{\mathrm{LGAD1}}$ and $\sigma_{\mathrm{LGAD2}}$):
        \begin{equation}
            \sigma_{\Delta T_{\mathrm{ref}}}^2 = \sigma_{\mathrm{LGAD1}}^2 + \sigma_{\mathrm{LGAD2}}^2
            \label{sigma_delta_T}
        \end{equation}
        Next, the timestamp of the diamond sensor was compared against the mean ToA of both LGADs:
        \begin{equation}
            \Delta T_{\mathrm{Diamond\text{-}Ref}} = \mathrm{ToA}_{\mathrm{Diamond}} - \frac{\mathrm{ToA}_{\mathrm{LGAD1}} + \mathrm{ToA}_{\mathrm{LGAD2}}}{2}
            \label{delta_tdiff}
        \end{equation}
        The variance of this total time difference ($\sigma_{\mathrm{Total}}^2$) is given by:
        \begin{equation}
            \sigma_{\mathrm{Total}}^2 = \sigma_{\mathrm{Diamond}}^2 + \frac{\sigma_{\mathrm{LGAD1}}^2 + \sigma_{\mathrm{LGAD2}}^2}{4}
            \label{sigma_total}
        \end{equation}
        Substituting the direct measurement of the reference variance sum ($\sigma_{\Delta T_{\mathrm{ref}}}^2 = \sigma_{\mathrm{LGAD1}}^2 + \sigma_{\mathrm{LGAD2}}^2$) into Eq.~(\ref{sigma_total}) yields:
        \begin{equation}
            \sigma_{\mathrm{Diamond}} = \sqrt{\sigma_{\mathrm{Total}}^2 - \frac{\sigma_{\Delta T_{\mathrm{ref}}}^2}{4}}
            \label{sigma_diam}
        \end{equation}
        
        This formulation avoids any assumption regarding equal performance between LGAD$_1$ and LGAD$_2$, allowing the intrinsic diamond timing precision to be rigorously extracted solely from measurable distribution widths.
        
        In order to verify that the time-walk corrections were properly applied and to determine the timing precision, the ToA differences were plotted as a function of the ToT of the corresponding reference sensor. An example histogram obtained for diamond pad \#21 using the $^{12}$C beam at 402~$A$MeV is shown in Fig.~\ref{fig:c_totvstdiff_corr}, where we see a flat distribution that is ToT-independent. A projection onto the Y-axis ($\Delta\text{T}$) was performed across the entire ToT range and the resulting time difference distribution was fitted with a Gaussian function, shown in Fig.~\ref{fig:c_tdiff_corr}. The standard deviation obtained from the fit is $\sigma_{\mathrm{Total}} = 61.8$~ps. After extracting $\sigma_{\Delta T_{\mathrm{ref}}}$ in a similar way from the time difference plots between LGADs, $\sigma_{\mathrm{Diamond}}$ was calculated for each pad using Eq.~(\ref{sigma_diam}).
        
        Fig.~\ref{fig:he_sigma} shows the resulting $\sigma_{\text{Diamond}}$ distribution across all diamond pads for the 73~$A$MeV $^{4}$He beam. The overall timing precision averages approximately 100~ps RMS, with the two most damaged pads, \#15 and \#21, exhibiting a precision of approximately 160 and 200~ps RMS, respectively. The periodic peak structure repeating every six pads is consistent with the efficiency pattern shown in Fig.~\ref{fig:eff_maps} and confirms a direct correlation with the radiation-damaged region. For the 402~$A$MeV$^{12}$C beam, the results presented in Fig.~\ref{fig:c_sigma} yield a timing precision of 30 to 50~ps RMS across all pads, except for the most damaged pad \#21, which reaches 58~ps RMS.
        
    \subsection{Cluster Size Dependence on Beam Energy and Radiation Damage}
        The cluster size is an important performance parameter of the CBM T$_0$ detector system, as it directly reflects the interplay between the detector response, front-end electronics, and threshold settings. The system is required to operate over an exceptionally wide dynamic range, from light ions such as carbon to heavy ions such as gold, while maintaining excellent time precision, high detection efficiency, and high-rate capability. In addition, the detector response changes with accumulated radiation damage, resulting in a gradual reduction of the signal amplitude. Consequently, the cluster size provides a sensitive measure of the detector and readout performance under varying operating conditions. During detector commissioning, cluster size measurements serve as one of the primary observables for optimizing the discriminator thresholds and ensuring stable operation throughout the experiment.
        
\begin{figure}[b!]
            \centering
            \begin{subfigure}[c]{0.45\columnwidth}
            \centering
            \caption{\label{fig:he_73_cs} 73~$A$MeV $^{4}$He}
            \includegraphics[width=\linewidth, trim=0 0 0 1cm, clip]{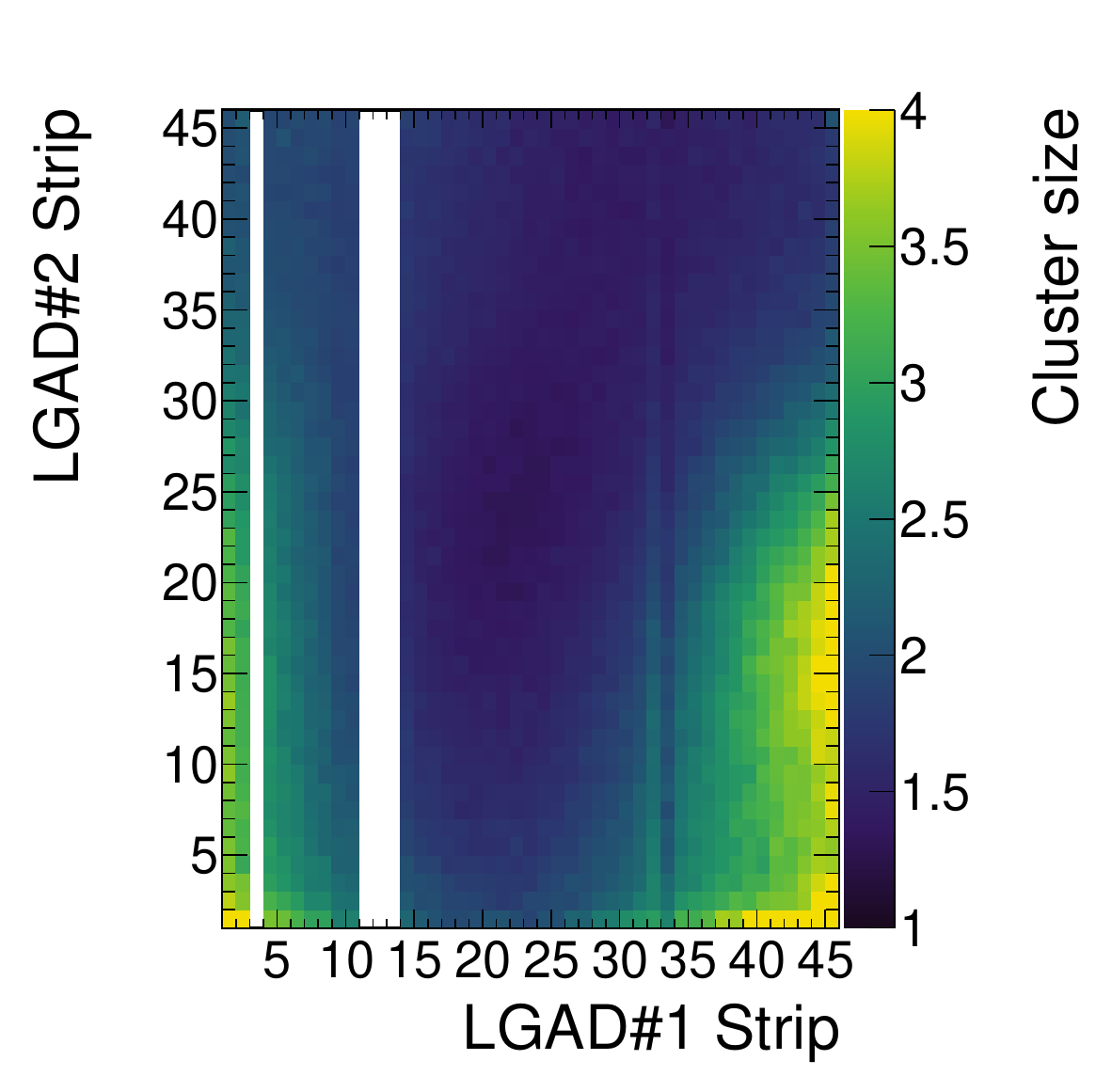}
            \end{subfigure}
            %\hfill
            \begin{subfigure}[c]{0.45\columnwidth}
            \centering
            \caption{\label{fig:c_402_cs} 402~$A$MeV $^{12}$C}
            \includegraphics[width=\linewidth, trim=0 0 0 1cm, clip]{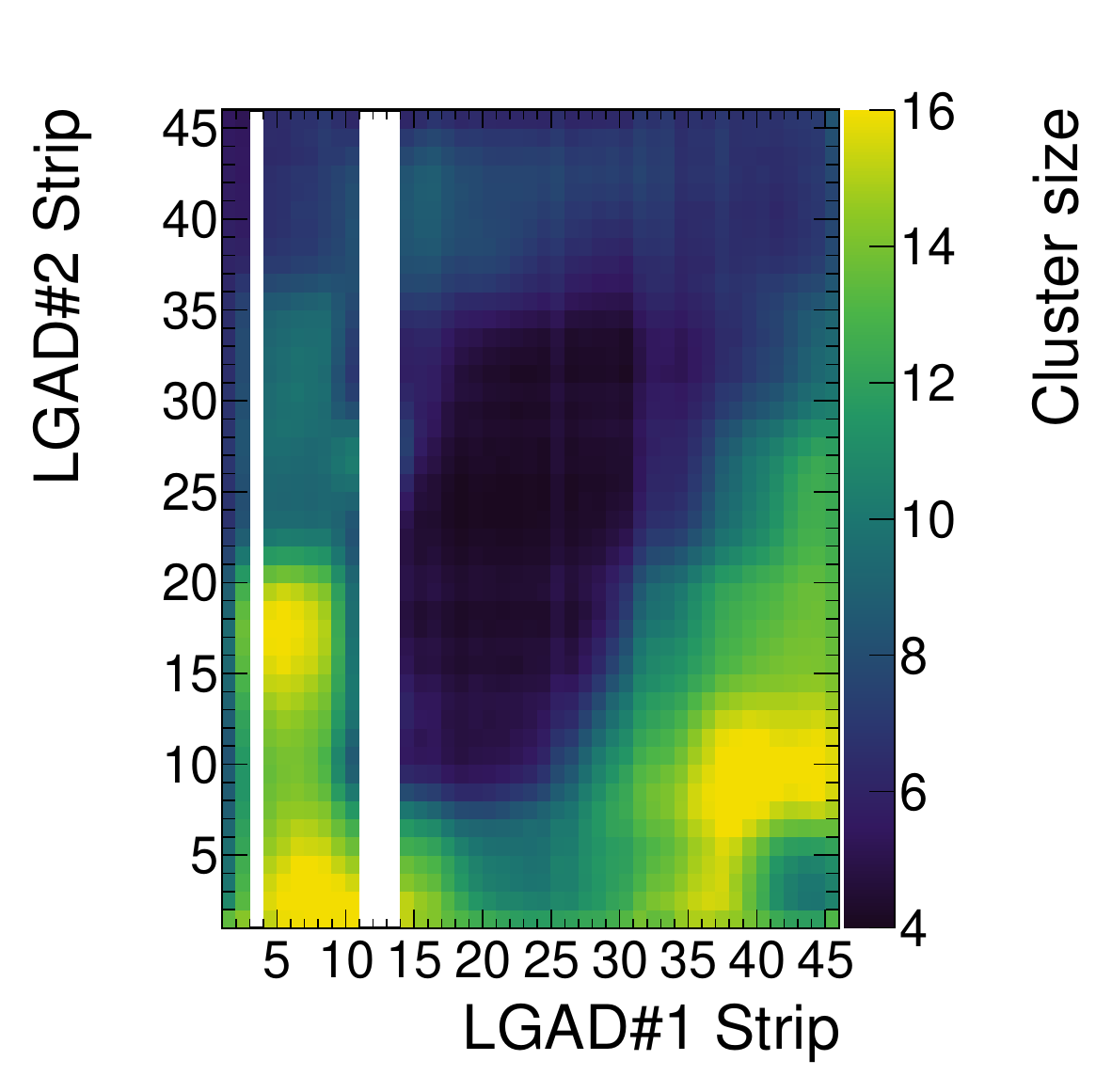}
            \end{subfigure}
            \caption{Cluster size maps for different beam particles. The Z axis scale is adjusted to improve the contrast of each individual histogram rather than for direct comparisons, with the highest cluster size shown for $^4$He being the lowest for $^{12}$C.}
            \label{fig:cs_maps}
        \end{figure}

        In this performance study of the CBM T$_0$ detector system, the discriminator thresholds were optimized based on the noise rate measured in the TDC, i.e. they were set as low as possible while maintaining stable operation in order to detect particles with the lowest expected $dE/dx$. The cluster size maps were calculated in a manner analogous to the efficiency maps by selecting time-correlated cluster centers in all three detectors. In addition, each selected hit was weighted by its associated cluster size. The resulting maps are shown in Fig.~\ref{fig:cs_maps}.
        
        For the 73~$A$MeV $^4$He beam (Fig.~\ref{fig:he_73_cs}), the average cluster size is close to 1 in the central region of the sensor, while it increases to values of up to 4 near the lower left and lower right corners. This behaviour is consistent with the efficiency map, which reveals the radiation damage pattern (Fig.~\ref{fig:c_402_eff}), as well as with the degraded time precision observed in the central region of the sensor (Fig.~\ref{fig:he_sigma}). The cluster size map obtained with the 402~$A$MeV $^{12}$C beam (Fig.~\ref{fig:c_402_cs}) exhibits a very similar spatial distribution but with significantly larger average cluster sizes, ranging from approximately 4 in the radiation-damaged region to about 16 in the lower corners. The pronounced difference in cluster size between the $^4$He and $^{12}$C measurements confirms that similar discriminator threshold settings were used in both cases and that these thresholds were optimized for the lowest expected $dE/dx$. The pad-to-pad variation in cluster size, which is more pronounced for the $^{12}$C beam, is attributed to small differences in the discriminator thresholds. Such variations strongly affect the registration of low-amplitude signals originating from capacitive coupling and electronic crosstalk, thereby influencing the measured cluster size.
        
        A clear separation between the upper and lower parts of the sensor is also observed. This effect is related to the sensor bonding, as the three upper pad rows are connected by longer bond wires than the lower rows, as partially visible in Figs.~\ref{fig:pad_diamond} and \ref{fig:He_C_test_setup}. The longer bond wires introduce a larger series inductance, which slightly reduces the peak amplitude of fast detector signals by temporarily storing part of the signal energy in the magnetic field. As a consequence, the measured ToT values become smaller, making low-amplitude capacitive coupling and crosstalk signals less likely to exceed the discriminator threshold and resulting in smaller measured cluster sizes.
        
    \subsection{Signal Degradation Margin for CBM Au-Beam Operation}
        \label{deg}
        To evaluate the expected long-term performance of the pcCVD diamond-based CBM T$_0$ detector for ultra-heavy-ion beams, specifically $^{197}$Au at kinetic energies above 2~$A$GeV, the maximum tolerable signal-amplitude degradation was estimated based on the detector performance measured with a 402~$A$MeV $^{12}$C beam. To relate these measurements to the expected CBM operating conditions, the stopping power in diamond was calculated using LISE++ (v17.16.6, ATIMA 1.2, LS-theory) \cite{lise}. The obtained $dE/dx$ values are 4274~keV/\textmu m for $^{197}$Au ions at 2~$A$GeV and 34.1~keV/\textmu m for $^{12}$C ions at 402~$A$MeV. The corresponding ratio,
        \begin{equation}
        \frac{
            (dE/dx)_{^{197}\mathrm{Au},\,2\,A\mathrm{GeV}}
        }{
            (dE/dx)_{^{12}\mathrm{C},\,402\,A\mathrm{MeV}}
        }
        =
        \frac{4274~\mathrm{keV}/\text{\textmu m}}
             {34.1~\mathrm{keV}/\text{\textmu m}}
        \approx 125.
        \end{equation}
        indicates that the energy deposition by the Au beam is approximately 125 times larger than that of the carbon beam used in the present measurements.
        
        Assuming that the collected signal amplitude scales approximately with the deposited energy, the successful operation with the $^{12}$C beam provides an estimate of the detector's tolerance to radiation-induced signal loss under CBM Au-beam conditions. In this approximation, a reduction of the collected signal by a factor of up to approximately 125 would bring the signal generated by 2~$A$GeV $^{197}$Au ions to the level measured with 402~$A$MeV $^{12}$C ions. Since the detector still achieved the required T$_0$ performance at this signal level, with an efficiency close to 100\% and a time precision better than 50~ps RMS, these results indicate a substantial operational margin for radiation-damaged pcCVD diamond sensors under the expected CBM conditions.

\section{Conclusion}
    \label{conclusion}
    In this work, the performance of the complete CBM T$_0$ detector system was evaluated using light-ion beams. The system incorporates several developments aimed at improving its performance and extending the operational lifetime of the pcCVD diamond sensor based $\mathrm{T}_0$ system under the demanding irradiation conditions expected in CBM. These include the transition from strip to pad metallization, which provides a better match to the highly non-uniform radiation-damage profile, and the use of a high-gain pre-amplification stage to maintain sensitivity as the sensor response degrades with accumulated radiation dose. In addition, the dedicated DOGMA readout system provides flexible per-channel threshold adjustment together with fast threshold-search and scanning procedures, enabling the readout conditions to be adapted to the local response of individual sensor regions.
    
    The beam measurements demonstrate that the complete readout and data-analysis chain, including channel time alignment, ToT calibration, time-walk correction, and hit selection, is well understood and provides the performance required for the CBM T$_0$ detector. In particular, measurements with 402~$A$MeV $^{12}$C ions provide an estimate of the detector's tolerance to radiation-induced signal loss under CBM Au-beam conditions. Based on the calculated energy-loss ratio and assuming an approximately proportional scaling between deposited energy and collected signal amplitude, an operational margin corresponding to a signal reduction by a factor of approximately 125 is obtained for 2~$A$GeV $^{197}$Au ions.
    
    An important outcome of this study is the characterization of the cluster size and its dependence on the detector operating conditions. For CBM operation with heavy ions, where the deposited energy is substantially larger than in the light-ion measurements presented here, the discriminator thresholds can be set well above the electronic noise level. We therefore propose to optimize the thresholds primarily on the basis of the measured cluster size rather than the noise count rate. This approach provides a practical criterion for maintaining high detection efficiency while limiting the number of simultaneously responding pads.
    The mechanical integration concept required for operation of the T$_0$ sensor in vacuum was also developed. It combines dedicated extension PCBs with a PCB-based vacuum feedthrough and a vacuum-compatible manipulator, allowing the detector position to be adjusted by several centimeters with respect to the beam axis. This provides the flexibility required to compensate for beam-position changes under different CBM operating conditions while maintaining the electrical connections between the sensor and the external readout electronics.
    
    Overall, the results demonstrate the suitability of the developed pad-based pcCVD diamond detector, DOGMA readout, and associated mechanical integration concept for the CBM T$_0$ system. The same concept is also planned to be adopted for the CBM beam-halo monitoring system.

\section{Acknowledgements}
    \label{acknowledgements}
    This research was funded in part by the Austrian Science Fund (FWF) Erwin-Schr\"odinger Grant Nr. J 4762-N, GSI-TU Darmstadt F\&E, DFG GRK 2128, DFG IGRK 2891 and the European Union’s Horizon 2020 research and innovation programme under grant agreement no 824093 (STRONG2020), 871072 and FAIR Phase-0. The financial support of the Austrian Ministry of Education, Science, and Research is gratefully acknowledged for providing beam time and research infrastructure at MedAustron.

%% The Appendices part is started with the command \appendix;
%% appendix sections are then done as normal sections
%%\appendix
%%\section{Example Appendix Section}
%%\label{app1}

%%Appendix text.

%% For citations use: 
%%      ~\cite{<label>} ==> [1]

%%
%%Example citation, See~\cite{lamport94}.

%% If you have bib database file and want bibtex to generate the
%% bibitems, please use
%%
%%  \bibliographystyle{elsarticle-num} 
%%  \bibliography{<your bibdatabase>}

%% else use the following coding to input the bibitems directly in the
%% TeX file.

%% Refer following link for more details about bibliography and citations.
%% https://en.wikibooks.org/wiki/LaTeX/Bibliography_Management

\bibliographystyle{elsarticle-num-names}
\bibliography{references}

\end{document}